# Anomalous Self-assembly of Architecturally Semiflexible Block Copolymers

Shifeng Nian,[1] Zhouhao Fan[2], Guillaume Freychet,[7] Mikhail Zhernenkov,[7] Stefanie Redemann,[4,5,6], Li-Heng Cai[1,2,3,*]

[1]Soft Biomatter Laboratory, Department of Materials Science and Engineering, University of Virginia, Charlottesville, VA 22904, USA
[2]Department of Chemical Engineering, UVA
[3]Department of Biomedical Engineering, UVA
[4]Department of Molecular Physiology and Biological Physics, UVA
[5]Department of Cell Biology, UVA
[6]Center for Membrane and Cell Physiology, UVA
[7]National Synchrotron Light Source-II, Brookhaven National Laboratory, Upton, NY 11973, USA

*Correspondence to liheng.cai@virginia.edu
ORCID: 0000-0002-6806-0566

**This PDF file includes:**
    Main Text
    Figures 1 to 4
    Table 1

**Abstract:** Block copolymer self-assembly is a fundamental process in which incompatible blocks spontaneously form organized microstructures with broad practical applications. Classical understanding is that the domain spacing is limited by the contour length of the polymer backbone. Here, using a combination of molecular design, chemical synthesis, small/wide-angle X-ray scattering, transmission electron microscopy, and electron tomography, we discover that this molecular picture does not hold for architecturally semiflexible block copolymers. For strongly segregated linear-semiflexible bottlebrush-linear triblock copolymers, the size of the bottlebrush domain can be twice the bottlebrush backbone contour length. The mechanism of such anomalous self-assembly likely is that the interfacial repulsion between the incompatible blocks is large enough to pull a part of the linear end-blocks into the bottlebrush domain. This effectively increases the bottlebrush domain size. Moreover, the semiflexible bottlebrush widens the regime for cylinder morphology that is associated with the volume fraction of the end blocks $f_C^{SFB} \in (0.10, > 0.40)$. This window is much wider than that for flexible linear block copolymers, $f_C^F \in (0.14, 0.35)$, and that predicted by recent self-consistent field theory for linear-bottlebrush block copolymers of the same chemistry and molecular architecture. Our experimental findings reveal previously unrecognized mechanisms for the self-assembly of architecturally complex block copolymers.

**Introduction**

A block copolymer (BCP) consists of two or more chemically distinct polymeric blocks linked by covalent bonds. BCP self-assembly is a fundamental process in which the incompatible blocks spontaneously form organized microstructures, which have broad applications in many technologically important areas;[1,2] examples include thermoplastic elastomers,[3] templates for lithography,[4] porous structures for filtration and separation,[5,6] and drug carriers.[7] Critical to these diverse applications is a wide range of macroscopic properties, which are largely determined by the type and the characteristic length of the self-assembled microstructures. The characteristic length scale afforded by classical flexible linear block copolymers is intrinsically small, however. It is largely determined by the molecular weight (MW) of polymers and is often below ~100 nm.[8] Further increasing the MW inevitably forms entanglements,[9] which slow down polymer dynamics and thus lead to uncontrollable self-assembly.[4] This limitation can be circumvented using bottlebrush-based BCPs,[6] in which at least one block is a bottlebrush polymer with a linear backbone densely grafted by many side chains.[10] This strategy has been exploited to create photonic crystals with characteristic lengths comparable to visible light,[11] solvent-free polymer networks of extreme softness[12,13] mimicking 'watery' biological tissues[14] elastomers with an exceptional combination of softness and structural coloration,[15] and very recently, a reprocessable soft elastomer for additive manufacturing.[16] These properties and applications highlight the potential of bottlebrush based BCPs as an emerging platform for materials design and innovation.

Unlike a linear polymer whose flexibility is constant, a bottlebrush can be physically flexible, semiflexible, or rigid. For example, increasing the size and/or grafting density of side chains stiffens the bottlebrush polymer.[10] To minimize excluded volume interactions, stiff polymers are prone to form highly ordered structures, a phenomenon often seen in liquid crystals[17] and in rod-coil BCPs.[18,19] However, this understanding does not apply to bottlebrush-based BCPs. For instance, recently we experimentally discovered that at small volume fractions of end blocks, $f$<0.05, linear-bottlebrush-linear (LBBL) triblock copolymers self-assemble to spherical microstructures regardless of the bottlebrush flexibility.[13] These findings were confirmed by self-consistent field theory (SCFT) for the self-assembly of bottlebrush BCPs.[20] Moreover, although the bottlebrush as a whole is geometrically bulky, its linear side chains can rearrange to adapt to curved interfaces to alleviate packing frustrations, as demonstrated by emulsions[21] and micelles[22] stabilized by bottlebrush-like surfactant molecules. These geometric and physical complexities offer a large parameter space for material design, but also pose challenges in understanding the self-assembly of bottlebrush-based BCPs. Consequently, despite its fundamental and technological importance, the understanding of how the bottlebrush molecular architecture determines BCP self-assembly is far from complete.

Here, we systematically investigate the effects of composition on the self-assembly of LBBL triblock copolymers. We focus on an architecturally semiflexible bottlebrush (SFB) that dramatically differs from its flexible linear counterpart in geometrical bulkiness and physical flexibility. Using a combination of molecular design, polymer synthesis, dark-field transmission electron microscopy (TEM), electron tomography, and small/wide-angle X-ray scattering (SAXS/WAXS), we establish the phase diagram for strongly segregated linear-semiflexible bottlebrush-linear triblock copolymers. The window for cylinder morphology, $f_C^{SFB} \in (0.10, > 0.40)$, is much wider than that for flexible linear block copolymers, $f_C^F \in (0.14, 0.35)$, and that predicted by recent self-consistent field theory for bottlebrush based block copolymers of the same molecular architecture.[20] Remarkably, regardless of the type of microstructure, the size of the bottlebrush domain is always larger than the contour length of the bottlebrush backbone. Even more surprisingly, the ratio between the two length scales becomes abnormally large around 2 at high volume fractions, and this observation is reproducible. We propose that the mechanism of such anomalous self-assembly likely is that a part of the linear end-blocks is pulled into the bottlebrush domain. This provides additional space for the side chains of the bottlebrush to rearrange and effectively increases the bottlebrush domain size. These results provide qualitatively new insights into the self-assembly of architecturally complex block copolymers.

**Results and Discussion**

*Molecular design and synthesis*

The flexibility $\kappa$ of a polymer is defined as the ratio of its contour length $\mathcal{L}_{\max}$ to twice the persistence length $\ell_p$, $\kappa \equiv \mathcal{L}_{\max}/(2\ell_p)$, and for a semiflexible polymer $\kappa \approx 1$. To design a semiflexible bottlebrush, we start with determining the dependencies of $\ell_p$ and $\mathcal{L}_{\max}$ on the bottlebrush molecular architecture (**Fig. 1a**). In a bottlebrush, the side chains are densely grafted to a backbone polymer, occupying a cylindrical space surrounding the backbone. The cross-section of the cylinder is about the size, $R_{sc}$, of a side chain. Within such a cylindrical space, a side chain occupies a volume, $R_{sc}^2 \lambda$, that is the product of the cross-section area, $R_{sc}^2$, and the distance between two neighboring grafting sites, $\lambda$. Because of mass conservation, this volume is equal to the volume of a side chain itself, $N_{sc} v_0$, in which $N_{sc}$ is the number of Kuhn monomers per side chain and $v_0$ is the volume of a Kuhn monomer. Therefore, the cross-section of the bottlebrush is $R_{sc} \approx (N_{sc} v_0/\lambda)^{1/2}$. The persistence length of the bottlebrush polymer is about its cross-section size,

$$\ell_p \approx R_{sc} \approx (N_{sc}v_0/\lambda)^{1/2} \qquad (1)$$

which increases with the side chain MW by a power of 1/2. This physical picture was described in our previous work[12] and others.[10,23] By contrast, the contour length of the bottlebrush is independent of the side chain MW; rather, it is proportional to number of side chains per bottlebrush $n_{BB}$:

$$\mathcal{L}_{\max} = n_{BB}\lambda \qquad (2)$$

The bottlebrush is essentially a 'fat' linear polymer, whose end-to-end distance $R$ is described by the worm-like chain model:[9] $R^2 = 2\ell_p \mathcal{L}_{\max} - 2\ell_p^2\left(1 - \exp\left(-\frac{\mathcal{L}_{\max}}{\ell_p}\right)\right)$. Above scaling arguments omit pre-factors on the order of unity as confirmed by molecular dynamics simulations of bottlebrush molecules.[10]

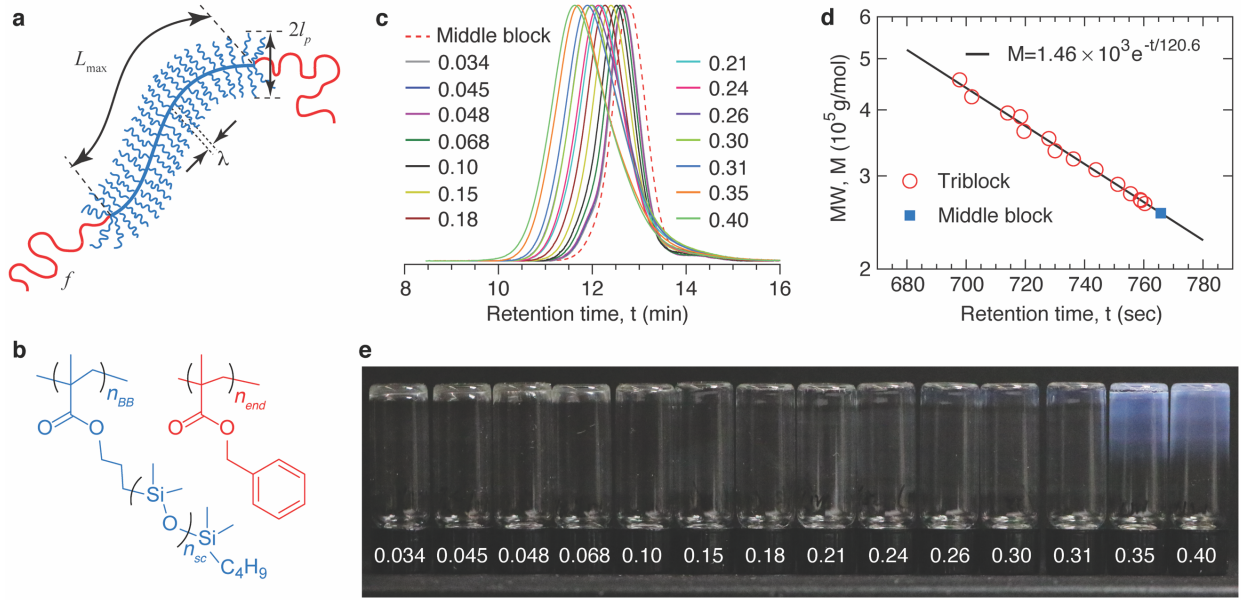

**Figure 1. Molecular design and synthesis of flexible-semiflexible bottlebrush-flexible (F-SFB-F) triblock copolymers. (a)** Schematic of a F-SFB-F triblock copolymer. The flexibility of a bottlebrush polymer is defined as $\kappa \equiv \mathcal{L}_{\max}/2\ell_p$, in which $\ell_p$ (eq. 1) and $\mathcal{L}_{\max}$ (eq. 2) are respectively the persistence length and the contour length of the bottlebrush. For a semiflexible

bottlebrush, $\kappa \approx 1$. $\lambda$ is the distance between two neighboring grafted side chains, and $f$ is the volume fraction of the end linear blocks. **(b)** In our model system, the side chain of the middle bottlebrush block is a linear polydimethylsiloxane (PDMS) with a molecular weight of 5000 g/mol, whereas the end blocks are linear poly(benzyl methacrylate) (PBnMA). The number of side chains per bottlebrush is fixed at $n_{BB} = 51$, whereas the number of repeating units $n_{end}$ of the end linear blocks is increased from 31 to 570. This results in triblock copolymers that have the same semiflexible bottlebrush middle block of contour length $\mathcal{L}_{max} = 12.8\ nm$ and persistence length $\ell_p \approx 5.8\ nm$, whereas the mass fraction $f_m$ of end blocks increases from 0.041 to 0.44. The corresponding volume fraction ranges from 0.034 to 0.40 based on the relation $f = f_m/[f_m + (1-f_m)(\rho_{PBnMA}/\rho_{PDMS})]$, in which $\rho_{PBnMA} = 1.18\ g/cm^3$ and $\rho_{PDMS} = 0.98\ g/cm^3$ are respectively the densities of PBnMA and PDMS. **(c)** GPC traces of the bottlebrush PDMS middle block (dashed line) and all F-SFB-F triblock copolymers (solid lines). **(d)** The logarithmic molecular weight, $M$, of all triblock copolymers decreases linearly with the increase of peak retention time, $t$: $M = 1.43 \times 10^8 g/mol\ \exp\left(-\frac{t}{120.6\ sec}\right)$. **(e)** At room temperature the polymers are optically transparent solid at $f < 0.26$ but gradually become blue at $f \geq 0.26$, and the blue color becomes brighter and more obvious at $f \geq 0.35$.

In our model system, the bottlebrush is formed by polymerizing methacrylate terminated poly(dimethyl siloxane) (MA-PDMS) macromonomers with a MW 5 kg/mol (**Fig. 1b**). This results in a bottlebrush of a grafting distance $\lambda = 0.25\ nm$. For a PDMS Kuhn monomer, the mass is $M_0 = 381$ g/mol, the length is $b = 1.3$ nm, and the volume is $v_0 = 6.50 \times 10^{-1}\ nm^3$. Using eq. (1), one obtains $\ell_p \approx 5.8$ nm. Therefore, it requires about $n_{BB} \approx 50$ side chains (eq. (2)) to create a semiflexible bottlebrush with $\kappa \approx 1$.

Guided by this molecular design, we extend our previously developed procedure[16] to synthesize flexible-semiflexible bottlebrush-flexible (F-SFB-F) triblock copolymers (**Materials and Methods**). The method is based on activators-regenerated-by-electron-transfer atom transfer radical polymerization (ARGET-ATRP),[24] and consists of two steps: first the synthesis of the middle bottlebrush block, and then the two end linear blocks (**Fig. S1**). In *Step I*, we use a bifunctional initiator, ethylene bis(2-bromoisobutyrate), to initiate the polymerization of MA-PDMS macromonomer. We closely monitor the reaction and stop the polymerization at a relatively low conversion 20%, up to which the conversion exhibits a linear dependence on reaction time (**Fig. S2**). This allows us to obtain a bottlebrush with precisely 51 side chains, which is determined based on the conversion of MA-PDMS macromonomers to a bottlebrush PDMS, as quantified by both proton nuclear magnetic resonance ($^1$H-NMR) spectroscopy (see **SI NMR spectra**) and gel permeation chromatography (GPC) (**Fig. S3**). Importantly, GPC reveals a narrowly distributed and symmetric retention profile with a small polydispersity index (PDI) equal to 1.15 (dashed line in **Fig. 1c**). These results demonstrate the controlled synthesis of a semiflexible PDMS bottlebrush.

In *Step II*, we use the bottlebrush as a di-functional macroinitiator to grow a linear poly(benzyl methacrylate) (PBnMA) to create triblock copolymers. For all polymers, we quantify the average MW of end blocks based on the conversion of BnMA monomers to PBnMA polymers as determined by $^1$H-NMR (see **SI NMR spectra**). The purified triblock copolymers are characterized using GPC (solid lines in **Fig. 1c**), based on which the PDI for all samples are determined (**Table 1**). As expected, the peak retention time decreases as the MW of the end blocks increases. Moreover, the MW determined from NMR decreases logarithmically with the increase of retention time (**Fig. 1d**). These results confirm the accuracy of MW of end blocks determined by NMR. Using the two-step synthesis, we obtain a series of PBnMA-bbPDMS-PBnMA triblock copolymers with the volume fraction $f$ of the end blocks ranging from 0.034 to 0.40. Importantly, all the triblock copolymers have the same semiflexible bottlebrush middle block with $L_{max} \approx 13$ nm and $\ell_p \approx 5.8$ nm, and only differ in the volume fraction of the end blocks. These polymers provide an ideal system for studying the self-assembly of F-SFB-F triblock copolymers.

At room temperature, all triblock copolymers are colorless except for those with high volume fractions, $f > 0.35$, which exhibit light blue color (**Fig. 1e**). This indicates ordered microstructures with characteristic lengths on the order of $\lambda/2n \approx 120 \, nm$, in which $\lambda = 380 \, nm$ is the wavelength of blue light and $n \approx 1.5$ is the refractive index of polymers.[8,25–27] This length scale is nearly ten times of the 13 nm contour length of the semiflexible bottlebrush backbone. Nevertheless, the transition from colorless to light blue indicates changes in microstructure as the volume fraction increases.

### *Determination of the type of microstructure*

The self-assembly of block copolymers is driven by the minimization of free energy.[28] As the molecular architecture becomes more complex, however, the process towards equilibrium can be trapped in metastable states.[4] And the probability of doing so becomes high for strongly segregated BCPs[29] like our system with a segregation strength $\chi N > 150$. Here, the Flory-Huggins interaction parameter $\chi$ between PBnMA and PDMS is about 0.26, the number of Kuhn monomers per triblock copolymer $N$ is greater than 650, and the effects of molecular architecture on the segregation strength are ignored. To this end, we use solvothermal annealing, a two-stage process consisting of a solvent vapor exposure followed by thermal annealing (**Materials and Methods**). During this process, the solvent molecules act as plasticizers and promote the mobility of individual polymer chains to achieve a thermodynamic equilibrium morphology.[30–32]

We use synchrotron source to perform SAXS/WAXS measurements to characterize the microstructure for the annealed polymers (**Materials and Methods**). In a typical measurement, an annealed polymer of cuboid shape with dimensions no smaller than 5×5×1 mm is mounted on a glass cover slip. Because the area of the sample is much larger than the beam size 250 μm × 20 μm, we perform measurements at multiple locations to ensure the consistency of the 2D SAXS patterns. An example scattering pattern of the polymer with $f = 0.35$ is shown in **Fig. 2a**. By subtracting the background from the cover slip and radially averaging the 2D pattern, we obtain 1D scattering intensity profile as a function of the magnitude of scattering wave vector $q$, as shown in **Fig. 2b**.

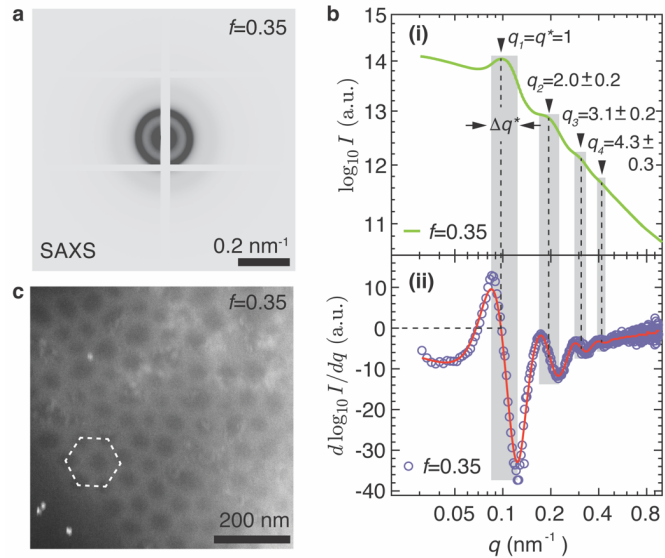

**Figure 2. SAXS and TEM determine the type of self-assembled microstructure.** (a) 2D SAXS pattern of sample $S_{13}$ with $f$=0.35 (**Table 1**). (b) (i) Radially averaged 1D scattering intensity $I$ as a function of the magnitude of wavevector $q$. (ii) The dependence of derivative $d\,(\log_{10} I)/dq$ on $q$. The solid red line is the spline fit to the data as the guidance for the eye. (c) A representative dark-field TEM image of a film with the thickness of 200 nm sliced using a microtome. The bright dots on the lower left are randomly deposited gold nanoparticles as fiducial markers for image alignment for electron tomography (**Movie S1**).

Next, we quantify the wavenumbers of all characteristic peaks. The primary peak is relatively sharp, exhibiting a local maximum that can be easily determined, as denoted by the first left arrow in **Fig. 2b**. Higher-order peaks are relatively broad, however; this makes it difficult to precisely determine their positions. Nevertheless, the scattered intensity $I$ changes rapidly with the wavenumber $q$ near a peak. This allows us to use the derivative, $d\,(\log_{10} I)/dq$, to locate the lower and the upper bounds of the peak, which are respectively associated with the local maximum and minimum, as illustrated by the shadowed regions in **Fig 2c**. A similar analysis is performed for all polymers (**Fig. S4**), and the upper and lower bounds of the primary peak are listed in **Table 1**. The average wavenumber, $q_n$, of the $n$-th peak corresponds to the deflection point between the lower bound and the upper bound, as denoted by the dashed lines in **Fig. 2c**.

**Table 1. Molecular parameters of F-SFB-F triblock polymers.** $M_{sc}$, molecular weight of side chains; $n_{BB}$, number of side chains per bottlebrush; $n_{end}$, number of chemical repeating units for each end linear PBnMA blocks; $f$, volume fraction of the end blocks; PDI, polydispersity index; $q^*$, wavenumber of the primary scattering peak; $d^* = 2\pi/q^*$, characteristic length associated with the primary scattering peak; $D$, domain diameter directly measured from TEM images; BCC S, body centered cubic sphere; Hex C, hexagonal cylinder; S-C, crossover between sphere and cylinder; NA, not applicable.

| Batch | Sample | Middle block | | | Triblock | | | Microstructure | | | | | Type |
| --- | --- | --- | --- | --- | --- | --- | --- | --- | --- | --- | --- | --- | --- |
| | | $M_{sc}$ (kDa) | $n_{BB}$ | PDI | $n_{end}$ | $f$ | PDI | $q^*$ (nm$^{-1}$) | | | $d^*$ (nm) | $D$ (nm) | |
| | | | | | | | | Peak | Upper | Lower | Peak | TEM | |
| 1 | S$_1$ | 5 | 51 | 1.15 | 31 | 0.034 | 1.17 | 0.3105 | 0.379 | 0.242 | 20.6 | NA | BCC S |
| | S$_2$ | | | | 41 | 0.045 | 1.17 | 0.310 | 0.364 | 0.256 | 20.8 | NA | BCC S |
| | S$_3$ | | | | 44 | 0.048 | 1.17 | 0.306 | 0.364 | 0.248 | 22.0 | 10.3±1.4 | BCC S |
| | S$_4$ | | | | 64 | 0.068 | 1.18 | 0.2615 | 0.302 | 0.221 | 25.5 | 10.4±1.3 | S-C |
| | S$_5$ | | | | 97 | 0.10 | 1.20 | 0.244 | 0.286 | 0.218 | 25.7 | 10.2±1.3 | S-C |
| | S$_6$ | | | | 151 | 0.15 | 1.31 | 0.198 | 0.243 | 0.178 | 31.7 | 16.9±1.9 | Hex C |

| | | | | | | | | | | | |
|---|---|---|---|---|---|---|---|---|---|---|---|
| | S$_7$ | | | 193 | 0.18 | 1.36 | 0.181 | 0.217 | 0.157 | 34.7 | 21.0±1.6 | Hex C |
| | S$_8$ | | | 226 | 0.21 | 1.54 | 0.165 | 0.196 | 0.148 | 38.1 | NA | Hex C |
| | S$_9$ | | | 278 | 0.24 | 1.55 | 0.159 | 0.194 | 0.136 | 39.5 | 22.6±2.4 | Hex C |
| | S$_{10}$ | | | 310 | 0.26 | 1.57 | 0.132 | 0.159 | 0.113 | 47.6 | NA | Hex C |
| | S$_{11}$ | | | 378 | 0.30 | 1.60 | 0.128 | 0.158 | 0.111 | 49.1 | NA | Hex C |
| | S$_{12}$ | | | 396 | 0.31 | 1.64 | 0.122 | 0.152 | 0.105 | 51.5 | NA | Hex C |
| | S$_{13}$ | | | 478 | 0.35 | 1.71 | 0.0994 | 0.1214 | 0.0855 | 63.2 | 45.4±5.4 | Hex C |
| | S$_{14}$ | | | 570 | 0.40 | 1.71 | 0.0911 | 0.113 | 0.072 | 68.9 | 53.8±9.2 | Hex C |
| 2 | S$_{15}$ | | 1.18 | 80 | 0.084 | 1.38 | 0.241 | 0.278 | 0.211 | 26.1 | NA | S-C |
| | S$_{16}$ | | | 371 | 0.30 | 1.57 | 0.131 | 0.164 | 0.114 | 47.9 | NA | Hex C |
| 3 | S$_{17}$ | | 1.20 | 186 | 0.18 | 1.58 | 0.174 | 0.204 | 0.152 | 36.1 | NA | Hex C |
| | S$_{18}$ | | | 275 | 0.24 | 1.60 | 0.156 | 0.188 | 0.128 | 40.3 | NA | Hex C |
| 4 | S$_{19}$ | | | 48 | 0.052 | 1.20 | 0.299 | 0.368 | 0.263 | 21.0 | NA | BCC S |
| | S$_{20}$ | | 1.17 | 124 | 0.12 | 1.26 | 0.218 | 0.259 | 0.192 | 28.8 | 12.6±1.1 | Hex C |
| | S$_{21}$ | | | 436 | 0.33 | 1.54 | 0.114 | 0.134 | 0.101 | 55.1 | NA | Hex C |
| 5 | S$_{22}$ | | 1.16 | 65 | 0.069 | 1.30 | 0.271 | 0.315 | 0.252 | 23.2 | NA | S-C |
| | S$_{23}$ | | | 600 | 0.41 | 1.71 | 0.0904 | 0.114 | 0.0691 | 69.5 | NA | Hex C |
| bbPDMS | | 5 | 51 | 1.15 | NA | NA | NA | 0.964 | 1.070 | 0.858 | 6.5 | NA | NA |

We attempt to determine the type of microstructure by comparing the positions of the characteristic peaks to the allowed reflections for a space group using ratios of $q/q^*$, where $q^*$ is the wavenumber of the primary scattering peak at the lowest wavenumber.[33] For example, the sample with $f = 0.35$ exhibits a relation $q/q^* = 1.0, 2.0 \pm 0.2, 3.1 \pm 0.2$, and $4.3 \pm 0.3$ (**Fig. 2c**) Within measurement error, unfortunately, this relation can be associated with either a hexagonal lattice of cylinder or a periodic one-dimensional (1D) structure of lamellae (see **SI text**). Consequently, because of relatively broad higher-order peaks, the relation $q/q^*$ alone is insufficient for precisely determining the type of microstructure.

To complement SAXS measurements, we use TEM to directly visualize the morphology of the self-assembled polymers. For the polymer with $f = 0.35$, we use microtome to slice the same bulk polymer as used for SAXS to obtain a thin film with a thickness of 200 nm (**Materials and**

**Methods**). Instead of using conventional bright-field TEM, we develop a technique based on dark-field TEM that uses diffracted rather than transmitted beam to image microdomains; this allows a sharp contrast of different domains without staining (**Materials and Methods**). The observed two-dimensional (2D) morphology exhibits a hexagonal pattern (**Fig. 2d**), which is further confirmed by the three-dimensional (3D) microstructure rendered from electron tomography (**Movie S1**). Collectively, these results demonstrate that the polymer with $f = 0.35$ forms a cylinder rather than a lamellar microstructure. Importantly, these results demonstrate that the combination of SAXS and TEM allows for unambiguous determination of the type of microstructure.

### *Phase diagram*

Using a combination of SAXS and TEM, we determine the crossover volume fractions between different phases (**Fig. 3**). At relatively small volume fractions, $0.07 \leq f \leq 0.10$, the diffraction peaks exhibit relative positions, $q/q^* = \sqrt{1}, \sqrt{3}, \sqrt{7}$, suggesting a hexagonal lattice of cylinder (**Fig. 3a**). However, dark-field TEM images do not exhibit an obvious hexagonal pattern (**Fig. S5a, b**); this is likely because of relatively small domain sizes and weak contrast associated with low volume fractions of end blocks. Yet, at a slightly higher volume fraction, $f = 0.12$, TEM reveals a hexagonal cylinder morphology (**Fig. S5c**). Such an ordering becomes more obvious and is of long range for $f = 0.15$ (**Fig. 3b-i**). Therefore, the crossover volume fraction between sphere and cylinder phases is likely within $0.70 \leq f_{SC}^{SFB} \leq 0.10$.

Interestingly, even at a high volume fraction, $f = 0.40$, the microstructure remains to be cylinder, as clearly shown by the TEM image in **Fig. 3b-iv** and large area view in **Fig. S5d**. Determining the crossover volume fraction between cylinder and lamellae requires controlled synthesis of high-quality samples with $f > 0.40$, or polymers with the number of repeating units >1200. This represents a challenge for ATRP and will be the subject of future study. Nevertheless, these results show that the cylinder morphology continues at volume fractions no less than 0.40. The window for cylinder phase, $f_C^{SFB} \in (0.10, > 0.40)$, is much wider than that of strongly segregated linear flexible diblock and triblock copolymers, $f_C^F \in (0.14, 0.35)$.[34]

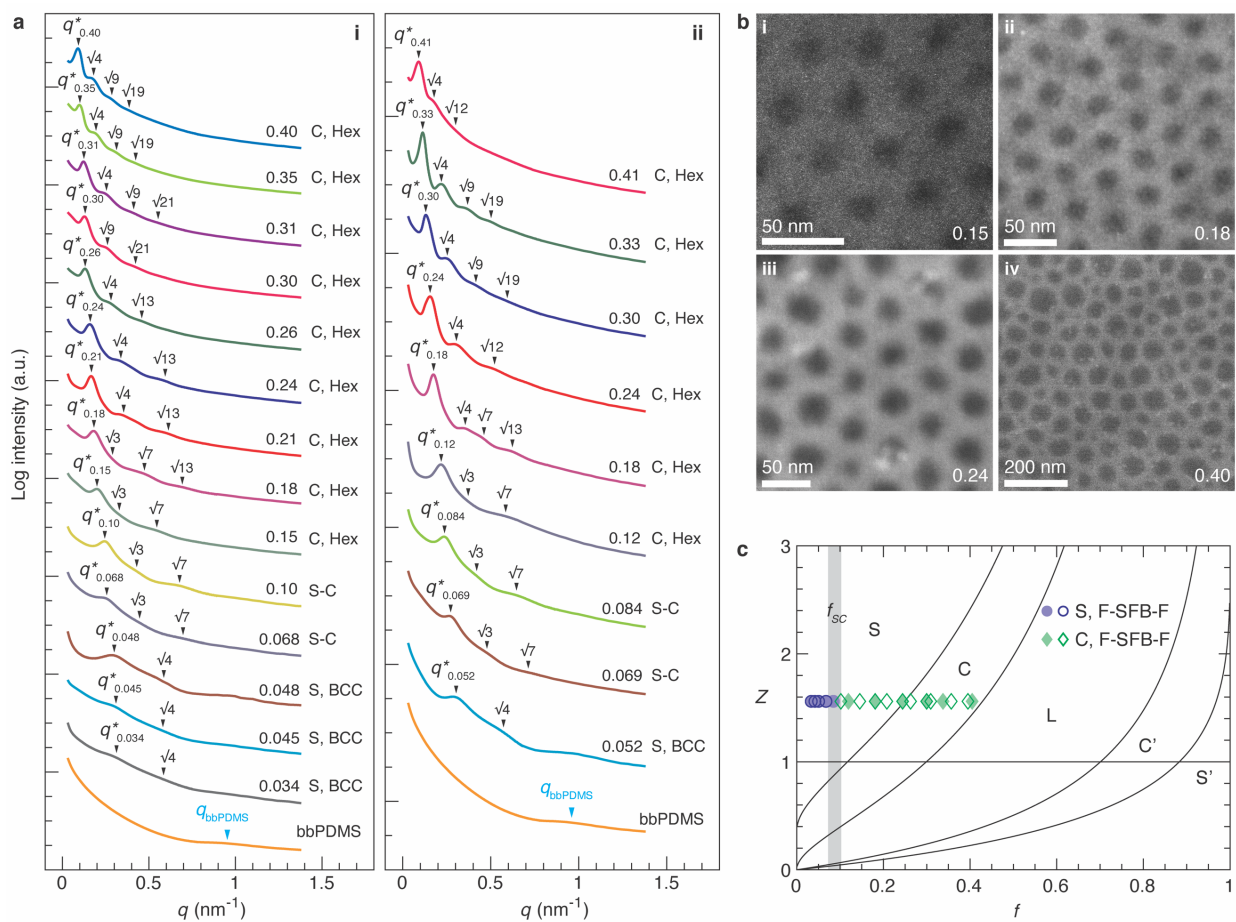

**Figure 3. Semiflexible bottlebrush widens the regime for cylinder morphology. (a)** Radially averaged 1D scattering intensity $I$ as a function of $q$: (i) polymers with the same bottlebrush middle block (batch 1, **Table 1**); (ii) polymers with the bottlebrush middle block synthesized from different batches (batches 2-5, **Table 1** and **Fig. S8**). The assignment of peak relations is based on the peak analysis described in **Fig. 2** and listed in **Fig. S4**. As $f$ increases, the primary scattering peak, $q^*$, shifts to smaller wavenumbers associated with larger characteristic length scales. The melt of bottlebrush PDMS exhibits a characteristic peak at $q_{bbPDMS} = 0.964$ nm$^{-1}$. The associated length scale, $2\pi/q_{bbPDMS} = 6.5$ nm, corresponds to the distance between the backbone of two neighboring bottlebrush molecules. This value is smaller than twice the size of the side chain, 11.6 nm, suggesting interdigitated side chains between neighboring bottlebrush molecules, as illustrated in **Fig. 4g**. **(b)** Representative dark-field TEM images of the self-assembled microstructures. Dark regions are domains formed by end linear PBnMA blocks, and white regions are bottlebrush PDMS domains. **(c)** A two-parameter $(\mathcal{Z}, f)$ phase diagram predicted by a recent SCFT study for the self-assembly of strongly segregated bottlebrush-based AB-type BCPs [35]. $\mathcal{Z}$ is determined by the difference in molecular architecture between the A and the B block: $\mathcal{Z} \equiv \beta(l_A b_A/v_A)^{1/2}(l_B b_B n_B)^{1/4}$. Here, $l$ and $v$ are respectively the size and the volume of a chemical monomer, $b$ is the length of a Kuhn segment, $n$ is the number of chemical monomers per side chain, and $\beta$ is a numerical factor on the order of unity; the subscripts $A$ and $B$ respectively refer to the end linear blocks and the side chains of the bottlebrush. In the theoretical study, a $\mathcal{Z}$ value equals to 1.6 is used for the bottlebrush with the same chemistry and molecular architecture as in this current study. Solid lines are predicted boundaries between different types of microstructure: S and S' – sphere; C and C' – cylinder; L – lamellae. Empty blue and green symbols are experimental data obtained from F-SFB-F polymers with the bottlebrush synthesized from the same batch, and the filled symbols are those with the bottlebrush synthesized from different batches. Circles: BCC packing of spheres; diamonds: hexagonal lattice of cylinders. The shadowed region indicates the crossover between sphere and cylinder phases.

Surprisingly, our experimental findings contradict recent SCFT theoretical predictions for bottlebrush-based BCPs.[35] The theoretical study uses bottlebrush polymers with the same chemistry and molecular architecture as in our current study. This allows us to apply the theory directly to the F-SFB-F triblock polymers without introducing any additional assumptions. The predicted window for the cylinder phase, $f_C^{SCFT} \in (0.22, 0.45)$, is much narrower than our experimental findings, as shown by the colored symbols in **Fig. 3c**. Such a large discrepancy indicates that the existing SCFT calculation cannot explain our experimental findings.

The phase behavior for F-SFB-F triblock copolymers is dramatically different from that of semiflexible linear BCPs. Existing experimental studies for linear flexible-semiflexible-flexible (F-SF-F) triblock copolymers[36] reveal either a disordered phase or an ordered lamellar microstructure. Consistent with these experimental findings, SCFT calculations suggest that linear flexible-semiflexible (F-SF) diblock copolymers[37,38] are prone to form highly ordered lamellar microstructures, except in two-dimensional space where non-lamellar microstructures become possible.[39] By contrast, F-SFB-F triblock copolymers exhibit both spherical and cylindrical microstructures. Compared to linear F-SF-F triblock copolymers, which transition from cylinder to lamellae morphology with $f$ above 0.20,[36] F-SFB-F triblock copolymers remain to be cylinder up to at least 0.40. These differences suggest that the self-assembly of architecturally semiflexible BCPs is both qualitatively and quantitatively different from that of simple linear semiflexible-flexible BCPs.

*Anomalous characteristic length scales*

The difference between architecturally semiflexible and simple linear block copolymers is further highlighted by the characteristic lengths of the microstructures. Based on the magnitude of the primary scattering wavevector $q^*$, we obtain the length scale, $d^* = 2\pi/q^*$, which corresponds to the first permitted Bragg peak of a microstructure, as respectively illustrated for BCC sphere, hexagonal cylinder, and lamellar phase in **Figs. 4a-c**. As the microstructure transitions from sphere to cylinder, $d^*$ increases dramatically from 20 nm to 70 nm. The relatively large $d^* \approx 70 \pm 20$ nm for the cylinder structure is consistent with the light blue color of the polymer (**Fig. 1d**).

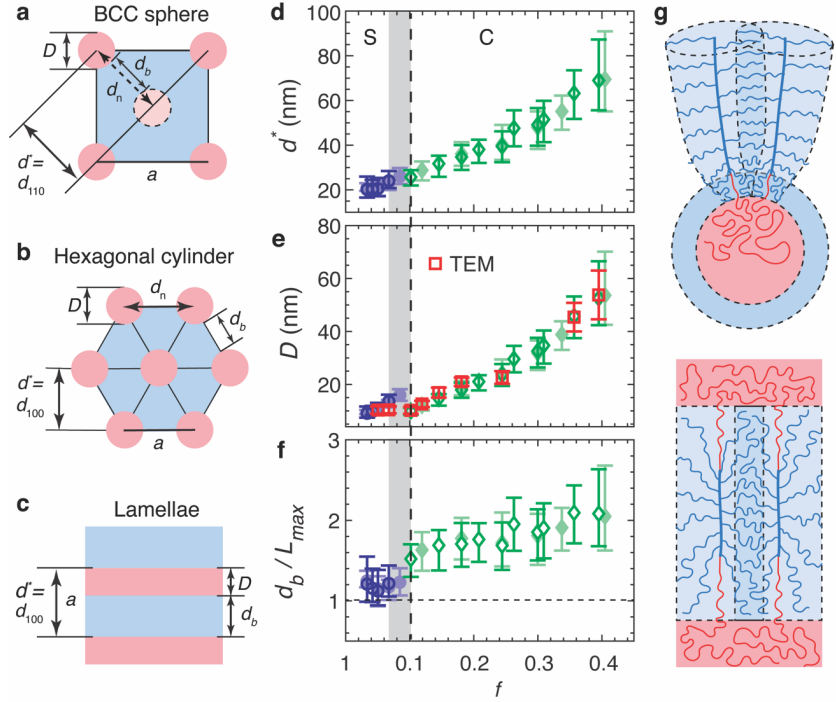

**Figure 4. Architecturally semiflexible bottlebrush rearranges its linear side chains to form remarkably large domains. (a-c)** Illustrations of characteristic lengths for BCC packing of spheres, hexagonal lattice of cylinders, and lamellar microstructures. $d^*$ corresponds to the first permitted Bragg peak of a microstructure, $a$ is the lattice constant, $D$ is the size of the domain formed by end blocks, $d_b$ is the distance bridging two nearest neighboring PBnMA domains, $d_n$ is the distance between the center of two nearest neighboring PBnMA domains. For BCC, $a = \sqrt{2}d_{110} = \sqrt{2}d^*$; for hexagonal cylinder, $a = (2/\sqrt{3})d_{100} = (2/\sqrt{3})d^*$; for lamellae, $d^* = d_{100} = a$. **(d-f)** Dependencies of the primary characteristic length $d^*$, PBnMA domain size $D$, normalized bridging distance $d_b/\mathcal{L}_{max}$ on the volume fraction $f$ of end blocks. Empty symbols are data for F-SFB-F polymers with the same batch of bottlebrush middle block, and the filled symbols are for polymers with different batches of bottlebrush middle block. Red squares are the diameters of end block domains directly measured from TEM (**Fig. S4**). **(g)** Illustration for the molecular picture of the anomalous self-assembly of F-SFB-F triblock copolymers in sphere, hexagonal cylinder, and lamellar microstructures. The end linear blocks are pulled out to generate additional space for the long side chains, alleviating packing frustration near curved interfaces and effectively increasing the size of the bottlebrush domain.

To quantify the contribution of end blocks to characteristic lengths, we calculate the size of the domain formed by the end blocks, $D$, based on two experimentally measured parameters: the volume fraction $f$ and the length $d^*$. For a BCC packing of spheres, $D = (3/\pi)^{\frac{1}{3}} f^{\frac{1}{3}} a$, in which the lattice constant $a = \sqrt{2} d^*$ (**Fig. 4a**); for an hexagonal lattice of cylinders, $D = (6/\sqrt{3}\pi)^{\frac{1}{2}} f^{\frac{1}{2}} a$, in which $a = (2/\sqrt{3}) d^*$ (**Fig. 4b**); for a 1D packing of lamellae, $D = fa = fd^*$ (**Fig. 4c**). As expected, for each microstructure, $D$ increases with $f$, or the MW of the end blocks (**Fig. 4e**). Moreover, the calculated values of $D$ based on SAXS measurements agree with that directly measured from TEM, as shown by red squares in **Fig. 4e** and histograms in **Fig. S6**. In addition, the volume fraction $f$ determined by NMR is consistent with that calculated from TEM, as shown by histograms in **Fig. S7** and **Table S1**. These results confirm the classification of the type of microstructure self-assembled by F-SFB-F triblock copolymers (**Fig. 3**).

To quantify the size of the bottlebrush domain, we define $d_b$ as the distance bridging two nearest neighboring PBnMA domains, as illustrated for different types of microstructure in **Figs. 4a-c**. Specifically, for a BCC packing of sphere, $d_b = \sqrt{3}a/2 - D$ and $a = \sqrt{2}d^*$; for an hexagonal lattice of cylinders, $d_b = a - D$ and $a = (2/\sqrt{3})d^*$; for a 1D packing of lamellae, $d_b = a - D$ and $a = d^*$. Further, we introduce a parameter, $d_b/\mathcal{L}_{\max}$, that is defined as the ratio of the bridging distance to the contour length of the bottlebrush backbone; this parameter describes the extent to which the bottlebrush is stretched.

Remarkably, the value of $d_b/\mathcal{L}_{\max}$ is abnormally large regardless of the types of microstructures. Specifically, $d_b/\mathcal{L}_{\max}$ is about 1.2 in the sphere phase (blue symbols in **Fig. 4f**); this is consistent with that observed in our previous study, in which experiments were designed for studying the effects of bottlebrush flexibility on self-assembly of linear-bottlebrush-linear triblock copolymers.[13] Moreover, in the cylinder phase, $d_b/\mathcal{L}_{\max}$ increases from 1.4 to nearly 2 as $f$ increases from 0.10 to 0.35 and stays around this value at higher $f$ (**Fig. 4f**). This is extremely surprising because $\mathcal{L}_{\max}$ is the maximum to which the bottlebrush backbone can be stretched. Thus, it is expected that $d_b/\mathcal{L}_{\max}$ must not be larger than 1. This upper limit has been proven not only for all kinds of simple linear BCPs regardless of their flexibility but also for bottlebrush diblock[40] and triblock[41] copolymers with all the blocks of the same diameter. Although polydisperse diblock copolymers with a PDI of about 2 can have large domain distances,[42–44] in our study the bottlebrush polymer is of relatively narrow dispersity with a PDI of 1.15. Moreover, all triblock copolymers have the same bottlebrush middle block. These preclude the polymer dispersity from being the cause of the unexpectedly large bridging distance. Nevertheless, to confirm our counterintuitive findings, we synthesize additional series of F-SFB-F triblock copolymers. Instead of using the same semiflexible bottlebrush for all triblock copolymers, we synthesize multiple batches of bottlebrush polymers that have the same average number of side chains but slightly different dispersity (**Fig. S8** and **Table 1**). Results from independent experiments are consistent with each other, as shown by the SAXS measurements (**Fig. 3a-ii**), the phase diagram (filled symbols in **Fig. 3c**), and the characteristic length scales (filled symbols in **Figs. 4d-f**). These results demonstrate that the observed large bridging distance is valid and reproducible.

We propose a molecular mechanism to explain the anomalous assembly of F-SFB-F ABA triblock copolymers. In the self-assembled microstructure, the interfacial repulsion between the highly incompatible A and B microdomains generates tension along the polymer backbone. Compared to a classical linear polymer, the cross-section of a semiflexible bottlebrush is much larger, resulting in a much larger interfacial area per triblock copolymer. Consequently, the resulted interfacial repulsion may be large enough to pull out a part of the linear chains from the A domains. This effectively increases the size of the bottlebrush domain, or the bridging distance, as illustrated in **Fig. 4g**.

Conceptually, the proposed molecular picture can be rationalized in the context of free energy minimization. We consider the total free energy, $F_t$, of a single F-SFB-F triblock copolymer. It consists of two components: $F_t = F_{int} + F_{ent}$, in which $F_{int}$ is the interfacial free energy that is proportional to the total interfacial area $A_t$, and $F_{ent}$ is the configurational entropy of polymer chains. The interfacial area consists of two parts: $A_t = A_l + A_c$, in which $A_l$ is determined by the "invaded" linear segment that is pulled into the bottlebrush domain, and $A_c$ is determined by the area of contact between the bottlebrush and the domain formed by the end blocks. As the size of the "invaded" segment increases, $A_l$ increases; by contrast, $A_c$ must decrease to conserve the bottlebrush volume. Therefore, there must be an optimized size for the "invaded" linear segment to minimize the interfacial free energy.

Similarly, the entropic free energy consists of two parts: $F_{ent} = F_{bb} + F_{sc}$, in which $F_{bb}$ and $F_{sc}$ are respectively the configurational entropy of the triblock polymer backbone and the side chains in the bottlebrush. Because of strong steric repulsion among the overlapping side chains, the backbone of the semiflexible bottlebrush is already pre-stretched to its contour length before microphase separation. Therefore, $F_{bb}$ is dominated by the "invaded" linear segment and increases with the segment size. By contrast, as the size of the "invaded" segment increases, it generates more space near the two ends of the bottlebrush for the densely grafted side chains to occupy; this results in less steric repulsion among the side chains and thus the decrease of $F_{sc}$. Thus, there must be an optimized size for the "invaded" linear segment to minimize the entropic free energy.

Quantitatively, the concept of "invaded" linear polymer segment permits the remarkably large length scales in the self-assembled microstructures. For instance, the contour length of a PDMS side chain is about 20 nm, much larger than the length, ~7 nm, required for the largest bridging distance at high volume fractions, as shown by the green diamonds on the right in **Fig. 4f**. Moreover, the "invaded" linear segment provides additional space for the densely grafted side chains to rearrange to alleviate packing frustration, which becomes exacerbated near highly curved interfaces for sphere and cylindrical phases, as illustrated in **Fig. 4g**. This molecular picture is qualitatively different from the one proposed for spherical microstructures self-assembled by linear-bottlebrush-linear triblock copolymers, in which the side chains near the ends of the bottlebrush are stretched along the backbone in the direction away from and perpendicular to the highly curved interface.[45] Taken together, the concept of "invaded" linear segments not only allows for the minimization of free energy but also avoids packing frustration, which are the two governing mechanisms for the self-assembly of block copolymers.[46]

**Conclusions**

We have systematically investigated the self-assembly of strongly segregated linear-semiflexible bottlebrush-linear triblock copolymers. We discover that the window for cylinder morphology, $f_C^{SFB} \in (0.10, > 0.40)$, is much wider than that for flexible linear ABA triblock copolymers, $f_C^F \in (0.14, 0.35)$, and that predicted by recent SCFT calculations, $f_C^{SCFT} \in (0.22, 0.45)$, for bottlebrush BCPs of the same chemistry and molecular architecture.[35] Remarkably, regardless of the types of microstructures, the size of the bottlebrush domain $d_b$ is always larger than the contour length $\mathcal{L}_{max}$ of the bottlebrush backbone. Even more surprisingly, the ratio between the two length scales, $d_b/\mathcal{L}_{max}$, becomes extremely large around 2 at high volume fractions; this observation is reproducible and further supported by the structural color of the self-assembled polymers. The anomalous self-assembly is likely because that a part of the linear end blocks is pulled into the bottlebrush domain, which effectively increases the bottlebrush domain size. The concept of such "invaded" linear end blocks is qualitatively different from classical understanding of ABA triblock copolymers, where the bridging distance between two neighboring A domains can never be larger than the contour length of the middle B block. The observed anomalous self-assembly of architecturally semiflexible block copolymers calls for further theoretical studies. It also cautions the use of conventional definition of bottlebrush contour length as that of the bottlebrush backbone,[10,16,47,48] which often ignores the contribution of side chains. The side chains can rearrange to occupy the space near the ends of the bottlebrush backbone. This effectively increases the bottlebrush contour length, and such an increase becomes significant for semiflexible bottlebrush polymers with long side chains. Nevertheless, our experimental findings reveal previously unrecognized mechanisms for the self-assembly of architecturally complex block copolymers. This knowledge may inform the design of multifunctional microstructures with an exceptional combination of long-range ordering and large characteristic length scales.

**Experimental Section**

**Materials.** MCR-M17, monomethacryloxypropyl terminated polydimethylsiloxane, average molar mass 5000 g/mol, was purchased from Gelest and purified using basic aluminum oxide columns to remove inhibitors before use. Benzyl methacrylate (96%) was purchased from Sigma Aldrich and purified using basic aluminum oxide columns to remove inhibitors before use. Copper(II) chloride ($CuCl_2$, 99.999%), *tris*[2-(dimethylamino)ethyl]amine ($Me_6TREN$), ethylene bis(2-bromoisobutyrate) (2-BiB, 97%), Tin(II) 2-ethylhexanoate ($Sn(EH)_2$, 92.5 – 100%), anisole (≥99.7%) and *p*-xylene (≥99.7%) were purchased from Sigma Aldrich and used as received. Methanol (Certified ACS), diethyl ether (Certified ACS), dimethylformamide (DMF, Certified ACS), tetrahydrofuran (THF, Certified ACS) and THF (HPLC), were purchased from Fisher and used as received.

**Polymer synthesis and characterization.** To synthesize a F-SFB-F triblock copolymer, we first synthesize the middle bottlebrush block, and then use the bottlebrush as a macro-initiator to grow the end linear blocks (**Fig. S1**). Here, we describe the detailed synthesis protocol using the F-SFB-F polymer with the highest volume fraction as an example.

*Step I. Synthesis of bottlebrush poly(dimethylsiloxane).* A 100 mL Schlenk flask is charged with 2f-BiB (7.2 mg, 0.02 mmol), MCR-M17 (30 g, 6 mmol), *p*-xylene (10 mL) and anisole (10 mL). We dissolve $Me_6TREN$ (46 mg, 0.2 mmol) and $CuCl_2$ (2.7 mg, 0.02 mmol) in 1 mL DMF to make a catalyst solution. Then, we add 120 μL catalyst solution, containing $2.4 \times 10^{-2}$ mmol $Me_6TREN$ and $2.4 \times 10^{-3}$ mmol $CuCl_2$, to the mixture and bubble it with nitrogen for 60 min to remove oxygen. Afterwards, the reducing agent, $Sn(EH)_2$ (38.9 mg, 0.096 mmol) in 200 μL *p*-xylene, is quickly added to the reaction mixture using a glass syringe. We seal the flask and then immerse it in an oil bath at 60°C to start the reaction. We stop the reaction after 205 min and take a small amount of mixture to determine the conversion using proton NMR (see **SI NMR Spectra**) and GPC (**Fig. S3**). These are two methods that allow independent measurement of the conversion, and both give the same value 17%. This results in a bottlebrush polymer with the degree of polymerization 51, corresponding to the number average molecular weight of 255 kg/mol.

The rest reaction mixture is diluted with THF and passed through a neutral aluminum oxide column to remove the catalyst. The collected solution is concentrated by a rotary evaporator (Buchi R-205). To separate the bottlebrush polymer from the unreacted macromonomers, we create a co-solvent, a mixture of methanol and diethyl ether with a volume ratio 3:2, which is a good solvent for the macromonomers but not for the bottlebrush PDMS. After precipitation, we further centrifuge the mixture to separate the polymer from the solvent, re-dissolve the separated polymer in THF to make a homogenous solution. We repeat this precipitation procedure for seven times to ensure that all unreacted macromonomers and impurities are completely removed. We use GPC to measure the PDI of the final product, which is 1.15 for this bbPDMS (**Fig. 1c**). At room temperature, the bbPDMS is a viscous, transparent liquid.

*Step II. Synthesis of LBBL triblock copolymers.* A 25 mL Schlenk flask is charged with benzyl methacrylate (BnMA, 893 mg, 5.07 mmol), macroinitiator (bbPDMS, 255 kg/mol, 255 mg, $1\times10^{-3}$ mmol), *p*-xylene (2.6 mL) and anisole (1.3 mL). We dissolve Me$_6$TREN (46 mg, 0.2 mmol) and CuCl$_2$ (2.7 mg, 0.02 mmol) in 1 mL DMF to make a catalyst solution. We add 85 µL catalyst solution, containing $1.7\times10^{-2}$ mmol Me$_6$TREN and $1.7\times10^{-3}$ mmol CuCl$_2$, to the mixture and bubble it with nitrogen for 45 min to remove oxygen. Afterwards, reducing agent, Sn(EH)$_2$ (27.5 mg, 6.8×10$^{-2}$ mmol) in 200 μL *p*-xylene, is quickly added to the reaction mixture using a glass syringe. Then, we seal the flask and immerse it in an oil bath at 60°C. The reaction is stopped after 360 min. The reaction mixture is diluted in THF and passed through a neutral aluminum oxide column to remove the catalyst, and the collected solution is concentrated by a rotavapor. Instead of using a methanol-diethyl ether co-solvent as in *Step I*, we create a co-solvent, a mixture of methanol and acetone with a volume ratio 1:1 for precipitation for three times; this completely removes all unreacted monomers and almost all free PBnMA due to chain transfer reaction during ATRP. After purification, the sample is dried in a vacuum oven (Thermo Fisher, Model 6258) at room temperature for 24 h. A small amount of the polymer is used for $^1$H NMR analysis and GPC analysis. From $^1$H NMR, the degree of polymerization of BnMA is 1140, corresponding to a MW 100.4 kg/mol for each of the two end blocks. From GPC, the PDI is 1.71 for this triblock copolymer (**Fig. 1c**). At room temperature, the polymer is a semi-transparent, blue, stiff solid.

**$^1$H NMR characterization.** We use $^1$H NMR to determine the number of side chains per bottlebrush and the volume fraction of PBnMA. The former one is calculated based on the conversion of PDMS macromonomers to bottlebrush PDMS, which is measured by the NMR spectra of the raw reaction mixture, as documented in a previous publication.[13] The volume fraction of PBnMA is determined based on the NMR spectra of purified triblock copolymers (see **SI NMR Spectra**).

**GPC characterization.** GPC measurements are performed using TOSOH EcoSEC HLC-8320 GPC system with two TOSOH Bioscience TSKgel GMH$_{HR}$-M 5μm columns in series and a refractive index detector at 40°C. HPLC grade THF is used as the eluent with a flow rate of 1 mL/min. The samples are dissolved in THF with a concentration around 5 mg/mL. The GPC data of all bbPDMS polymers and the corresponding LBBL polymers are shown in **Fig. 1c** and **Fig. S8**. The molecular weight and PDI of all samples are summarized in **Table 1**.

**SAXS/WAXS measurements.** To prepare a sample for SAXS/WAXS characterization, we dissolve a triblock copolymer in toluene at a concentration of 100 mg/mL with a total volume of 3 mL in a glass vial and allow the solvent to slowly evaporate for 24 hours. Because toluene is a solvent close to be equally good for PBnMA and PDMS, it avoids the effects of solvent selectivity on the self-assembly. The structures of the sample do not change after being further thermal annealed for 6 hours in a vacuum oven at 180°C and then slowly cooled down at a rate of 0.5 °C/min to room temperature. For each sample, we use relatively large amount of polymer to obtain a bulk material with the smallest dimension larger than 1 mm, more than $10^4$ times the size of a triblock copolymer; this prevents substrate or boundary effects on self-assembly process.

We use the Soft Matter Interfaces (12-ID) beamline[49] at the Brookhaven National Laboratory to perform SAXS measurements on annealed bulk polymers. The sample-to-detector distance is 8.3 m and the radiation wavelength is $\lambda = 0.77$ Å. The scattered X-rays are recorded using an in-vacuum Pilatus 1M detector, consisting of 0.172 mm square pixels in a 941×1043 array. The raw SAXS images are converted into $q$-space, visualized in Xi-CAM software[50] and radially integrated using a custom Python code. The one-dimensional intensity profile, *I(q)*, is plotted as a function of the scattering wave vector, $|\vec{q}| = q = 4\pi\lambda^{-1}\sin(\theta/2)$, where $\theta$ is the scattering angle.

**Transmission electron microscopy.** We use a combination of pressurized solvent vapor and thermal annealing to prepare samples for TEM imaging. We build a glass chamber with a sample stage, on which is placed a carbon film coated copper TEM grid. Moreover, the bottom of the glass chamber is filled toluene. In parallel, we prepare a F-SFB-F polymer solution by dissolving the polymer in toluene with a concentration of 5 mg/mL. The polymer solution is passed through a syringe filter with pore size 0.45 μm to remove possible dusts. We add 7 μL polymer solution to the TEM grid, seal the glass chamber immediately, transfer the chamber into an oven with temperature of 100 °C, and maintain the temperature for 24 h. This generates a toluene vapor pressure of 75 kPa. Afterwards, the oven is slowly cooled down at a rate of 0.5 °C/min to room temperature. Further increasing the annealing time to 3 days does not change the morphology of the polymers.

We use hollow-cone dark-field TEM (FEI Titan) at the electron energy of 300 keV with a tilt angle of 0.805º to characterize the annealed samples. This tilt angle allows for sharp contrast between PDMS and PBnMA domains without staining. The size of spherical domains is calculated using ImageJ, and more than 200 domains are used to ensure sufficient statistics.

**Electron tomography.** We use microtome to slice the polymer for electron tomography. To prepare the sample for microtome, we mount a piece of annealed bulked polymer on a "dummy" block using a two-part adhesive epoxy glue (Epoxy MS-907 Plus, Miller-Stephenson). The glue is left to cure for 24h at room temperature. Using Reichert Leica Ultracut S Microtome (Leica Microsystems, Vienna, Austria) equipped with a diamond knife, we slice the sample to a serial of sections with 200 nm thickness each and collect the sections on formvar-coated copper slot grids.

Colloidal gold nanoparticles (15 nm; Sigma-Aldrich) are deposited to both sides of semi-thick sections collected on copper slot grids to serve as fiducial markers for subsequent image alignment.

For single-axis electron tomography, series of tilted views are recorded using a F20 electron microscopy (Thermo-Fisher, formerly FEI) operating at 200 kV. Images are captured every 1° over a ±60° range and a pixel size of 0.3731 nm or 0.7439 nm using a Tietz TVIPS XF416 camera (4k×4k). We use the IMOD software package (http://bio3d.colourado.edu/imod) that contains all the programs needed for calculating electron tomograms.[51] For image processing, the tilted views are aligned using the positions of the colloidal gold particles as fiducial markers. Tomograms are computed using the R-weighted back-projection algorithm.[52]

**Supporting information:** Discussion on characteristic SAXS peaks for different types of microstructures. GPC traces of all bbPDMS and LBBL polymers. $^1$H NMR spectra of selected bbPDMS and all LBBL polymers. Example figures of determination of characteristic peaks of SAXS measurements. Example TEM images of LBBL polymers. Selected figures of domain diameters and distances of LBBL polymers. Movie of electron tomography measurement for LBBL polymer with $f$=0.35.

**Acknowledgements:** We thank Timothy Lodge (University of Minnesota), Edwin Thomas (Texas A&M), Mark Matsen (University of Waterloo), and Russel Spencer (University of Waterloo) for enlightening discussions. L.H.C. acknowledges the support from NSF (CAREER DMR-1944625) and ACS Petroleum Research Fund (PRF) (6132047-DNI). This research used the SMI beamline (12-ID) of the National Synchrotron Light Source II, a U.S. Department of Energy (DOE) Office of Science User Facility operated for the DOE Office of Science by Brookhaven National Laboratory under contract no. DE-SC0012704.

**Author contributions:** L.H.C. conceived the study. L.H.C. and S.N. designed the research. S.N. and L.H.C. performed the research. S.R. performed TEM tomography characterization. S.N. and L.H.C. analyzed data. F.Z. helped with polymer synthesis. G.F. and M.Z. helped with SAXS/WAXS measurements and data analysis. L.H.C. and S.N. wrote the paper. All authors reviewed and commented on the paper. L.H.C. supervised the research.

**Competing interests:** The authors declare no competing interests.

**References**

(1) Bates, C. M.; Bates, F. S. 50th Anniversary Perspective: Block Polymers - Pure Potential. *Macromolecules* **2017**, *50* (1), 3–22. https://doi.org/10.1021/acs.macromol.6b02355.

(2) Lodge, T. P. Block Copolymers: Past Successes and Future Challenges. *Macromol. Chem. Phys.* **2003**, *204* (2), 265–273. https://doi.org/10.1002/macp.200290073.

(3) Spontak, R. J.; Patel, N. P. Thermoplastic Elastomers: Fundamentals and Applications. *Curr. Opin. Colloid Interface Sci.* **2000**, *5* (5–6), 333–340. https://doi.org/10.1016/s1359-0294(00)00070-4.

(4) Segalman, R. A. Patterning with Block Copolymer Thin Films. *Mater. Sci. Eng. R Reports* **2005**, *48* (6), 191–226. https://doi.org/10.1016/j.mser.2004.12.003.

(5) Hillmyer, M. A. Nanoporous Materials from Block Copolymer Precursors. In *Advances in Polymer Science*; 2005; Vol. 190, pp 137–181. https://doi.org/10.1007/12_002.

(6) Verduzco, R.; Li, X.; Pesek, S. L.; Stein, G. E. Structure, Function, Self-Assembly, and Applications of Bottlebrush Copolymers. *Chem. Soc. Rev.* **2015**, *44* (8), 2405–2420. https://doi.org/10.1039/c4cs00329b.

(7) Kataoka, K.; Harada, A.; Nagasaki, Y. Block Copolymer Micelles for Drug Delivery: Design, Characterization and Biological Significance. *Adv. Drug Deliv. Rev.* **2012**, *64*, 37–48. https://doi.org/10.1016/j.addr.2012.09.013.

(8) Edrington, B. A. C.; Urbas, A. M.; Derege, P.; Chen, C. X.; Swager, T. M.; Hadjichristidis, N.; Xenidou, M.; Fetters, L. J.; Joannopoulos, J. D.; Fink, Y.; Thomas, E. L. Polymer-Based Photonic Crystals. *Adv. Mater.* **2001**, *13* (6), 421–425.

(9) Rubinstein, M.; Colby, R. H. *Polymer Physics*; Oxford University Press: Oxford ; New York, 2003.

(10) Paturej, J. J.; Sheiko, S. S.; Panyukov, S.; Rubinstein, M. Molecular Structure of Bottlebrush Polymers in Melts. *Sci. Adv.* **2016**, *2* (November), e1601478. https://doi.org/10.1126/sciadv.1601478.

(11) Sveinbjörnsson, B. R.; Weitekamp, R. A.; Miyake, G. M.; Xia, Y.; Atwater, H. A.; Grubbs, R. H. Rapid Self-Assembly of Brush Block Copolymers to Photonic Crystals. *Proc. Natl. Acad. Sci.* **2012**, *109* (36), 14332–14336. https://doi.org/10.1073/pnas.1213055109.

(12) Cai, L. H.; Kodger, T. E.; Guerra, R. E.; Pegoraro, A. F.; Rubinstein, M.; Weitz, D. A. Soft Poly(Dimethylsiloxane) Elastomers from Architecture-Driven Entanglement Free Design. *Adv. Mater.* **2015**, *27* (35), 5132–5140. https://doi.org/10.1002/adma.201502771.

(13) Nian, S.; Lian, H.; Gong, Z.; Zhernenkov, M.; Qin, J.; Cai, L. Molecular Architecture Directs Linear–Bottlebrush–Linear Triblock Copolymers to Self-Assemble to Soft Reprocessable Elastomers. *ACS Macro Lett.* **2019**, *8* (11), 1528–1534. https://doi.org/10.1021/acsmacrolett.9b00721.

(14) Levental, I.; Georges, P. C.; Janmey, P. A. Soft Biological Materials and Their Impact on Cell Function. *Soft Matter* **2007**, *3* (3), 1–9. https://doi.org/10.1039/b610522j.

(15) Vatankhah-Varnosfaderani, M.; Keith, A. N.; Cong, Y.; Liang, H.; Rosenthal, M.; Sztucki, M.; Clair, C.; Magonov, S.; Ivanov, D. A.; Dobrynin, A. V; Sheiko, S. S. Chameleon-like Elastomers with Molecularly Encoded Strain-Adaptive Stiffening and Coloration. *Science* **2018**, *359* (6383), 1509–1513. https://doi.org/10.1126/science.aar5308.

(16) Nian, S.; Zhu, J.; Zhang, H.; Gong, Z.; Freychet, G.; Zhernenkov, M.; Xu, B.; Cai, L.-H. Three-Dimensional Printable, Extremely Soft, Stretchable, and Reversible Elastomers from Molecular Architecture-Directed Assembly. *Chem. Mater.* **2021**, *33* (7), 2436–2445. https://doi.org/10.1021/acs.chemmater.0c04659.

(17) Gennes, P. G. de.; Prost, J. *The Physics of Liquid Crystals*; Clarendon Press, 1993.

(18) Olsen, B. D.; Segalman, R. A. Self-Assembly of Rod–Coil Block Copolymers. *Mater. Sci. Eng. R Reports* **2008**, *62* (2), 37–66. https://doi.org/10.1016/J.MSER.2008.04.001.

(19) Lee, M.; Cho, B. K.; Zin, W. C. Supramolecular Structures from Rod-Coil Block Copolymers. *Chem. Rev.* **2001**, *101* (12), 3869–3892. https://doi.org/10.1021/cr0001131.

(20) Zhulina, E. B.; Sheiko, S. S.; Dobrynin, A. V.; Borisov, O. V. Microphase Segregation in the Melts of Bottlebrush Block Copolymers. *Macromolecules* **2020**, *53* (7), 2582−2593. https://doi.org/10.1021/acs.macromol.9b02485.

(21) Li, Y.; Zou, J.; Das, B. P.; Tsianou, M.; Cheng, C. Well-Defined Amphiphilic Double-Brush Copolymers and Their Performance as Emulsion Surfactants. *Macromolecules* **2012**, *45* (11), 4623–4629. https://doi.org/10.1021/ma300565j.

(22) Fenyves, R.; Schmutz, M.; Horner, I. J.; Bright, F. V.; Rzayev, J. Aqueous Self-Assembly of Giant Bottlebrush Block Copolymer Surfactants as Shape-Tunable Building Blocks. *J. Am. Chem. Soc.* **2014**, *136* (21), 7762–7770. https://doi.org/10.1021/ja503283r.

(23) Daniel, W. F. M.; Burdyńska, J.; Vatankhah-Varnoosfaderani, M.; Matyjaszewski, K.; Paturej, J.; Rubinstein, M.; Dobrynin, A. V.; Sheiko, S. S. Solvent-Free, Supersoft and Superelastic Bottlebrush Melts and Networks. *Nat. Mater.* **2016**, *15* (2), 183–189. https://doi.org/10.1038/nmat4508.

(24) Matyjaszewski, K.; Jakubowski, W.; Min, K.; Tang, W.; Huang, J.; Braunecker, W. A.; Tsarevsky, N. V. Diminishing Catalyst Concentration in Atom Transfer Radical Polymerization with Reducing Agents. *Proc. Natl. Acad. Sci.* **2006**, *103* (42), 15309–15314. https://doi.org/10.1073/pnas.0602675103.

(25) Paquet, C.; Kumacheva, E. Nanostructured Polymers for Photonics. *Mater. Today* **2008**, *11* (4), 48–56. https://doi.org/10.1016/S1369-7021(08)70056-7.

(26) Fink, Y.; Winn, J. N.; Fan, S.; Chen, C.; Michel, J.; Joannopoulos, J. D.; Thomas, E. L. A Dielectric Omnidirectional Reflector. *Science* **1998**, *282* (5394), 1679–1682. https://doi.org/10.1126/science.282.5394.1679.

(27) Urbas, A.; Sharp, R.; Fink, Y.; Thomas, E. L.; Xenidou, M.; Fetters, L. J. Tunable Block Copolymer/Homopolymer Photonic Crystals. *Adv. Mater.* **2000**, *12* (11), 812–814. https://doi.org/10.1002/(SICI)1521-4095(200006)12:11<812::AID-ADMA812>3.0.CO;2-8.

(28) Hu, H.; Gopinadhan, M.; Osuji, C. O. Directed Self-Assembly of Block Copolymers: A Tutorial Review of Strategies for Enabling Nanotechnology with Soft Matter. *Soft Matter* **2014**, *10* (22), 3867. https://doi.org/10.1039/c3sm52607k.

(29) Bates, F. S.; Fredrickson, G. H. Block Copolymer Thermodynamics: Theory and Experiment. *Annu. Rev. Phys. Chem.* **1990**, *41* (1), 525–557. https://doi.org/10.1146/annurev.pc.41.100190.002521.

(30) Kim, S. H.; Misner, M. J.; Xu, T.; Kimura, M.; Russell, T. P. Highly Oriented and Ordered Arrays from Block Copolymers via Solvent Evaporation. *Adv. Mater.* **2004**, *16* (3), 226–231. https://doi.org/10.1002/adma.200304906.

(31) Gotrik, K. W.; Ross, C. A. Solvothermal Annealing of Block Copolymer Thin Films. *Nano Lett.* **2013**, *13* (11), 5117–5122. https://doi.org/10.1021/nl4021683.

(32) Sinturel, C.; Vayer, M.; Morris, M.; Hillmyer, M. A. Solvent Vapor Annealing of Block Polymer Thin Films. *Macromolecules* **2013**, *46* (14), 5399–5415.

https://doi.org/10.1021/ma400735a.
(33) Hamley, I. W.; Castelletto, V. Small-Angle Scattering of Block Copolymers in the Melt, Solution and Crystal States. *Prog. Polym. Sci.* **2004**, *29* (9), 909–948. https://doi.org/10.1016/j.progpolymsci.2004.06.001.
(34) Matsen, M. W. Equilibrium Behavior of Asymmetric ABA Triblock Copolymer Melts. *J. Chem. Phys.* **2000**, *113* (13), 5539–5544. https://doi.org/10.1063/1.1289889.
(35) Shi, W.; Lynd, N. A.; Montarnal, D.; Luo, Y.; Fredrickson, G. H.; Kramer, E. J.; Ntaras, C.; Avgeropoulos, A.; Hexemer, A. Toward Strong Thermoplastic Elastomers with Asymmetric Miktoarm Block Copolymer Architectures. *Macromolecules* **2014**, *47* (6), 2037–2043. https://doi.org/10.1021/ma402566g.
(36) Koga, M.; Abe, K.; Sato, K.; Koki, J.; Kang, S.; Sakajiri, K.; Watanabe, J.; Tokita, M. Self-Assembly of Flexible-Semiflexible-Flexible Triblock Copolymers. *Macromolecules* **2014**, *47* (13), 4438–4444. https://doi.org/10.1021/ma500798z.
(37) Kumar, N. A.; Ganesan, V. Communication: Self-Assembly of Semiflexible-Flexible Block Copolymers. *J. Chem. Phys.* **2012**, *136* (10), 5057. https://doi.org/10.1063/1.3692601.
(38) Song, W.; Tang, P.; Qiu, F.; Yang, Y.; Shi, A.-C. C. Phase Behavior of Semiflexible-Coil Diblock Copolymers: A Hybrid Numerical SCFT Approach. *Soft Matter* **2011**, *7* (3), 929–938. https://doi.org/10.1039/c0sm00841a.
(39) Gao, J.; Tang, P.; Yang, Y. Non-Lamellae Structures of Coil-Semiflexible Diblock Copolymers. *Soft Matter* **2013**, *9* (1), 69–81. https://doi.org/10.1039/c2sm26758f.
(40) Dalsin, S. J.; Rions-Maehren, T. G.; Beam, M. D.; Bates, F. S.; Hillmyer, M. A.; Matsen, M. W. Bottlebrush Block Polymers: Quantitative Theory and Experiments. *ACS Nano* **2015**, *9* (12), 12233–12245. https://doi.org/10.1021/acsnano.5b05473.
(41) Sunday, D. F.; Chang, A. B.; Liman, C. D.; Gann, E.; Delongchamp, D. M.; Thomsen, L.; Matsen, M. W.; Grubbs, R. H.; Soles, C. L. Self-Assembly of ABC Bottlebrush Triblock Terpolymers with Evidence for Looped Backbone Conformations. *Macromolecules* **2018**, *51* (18), 7178–7185. https://doi.org/10.1021/acs.macromol.8b01370.
(42) Lynd, N. A.; Hillmyer, M. A. Influence of Polydispersity on the Self-Assembly of Diblock Copolymers. *Macromolecules* **2005**, *38* (21), 8803–8810. https://doi.org/10.1021/ma051025r.
(43) Hustad, P. D.; Marchand, G. R.; Garcia-Meitin, E. I.; Roberts, P. L.; Weinhold, J. D. Photonic Polyethylene from Self-Assembled Mesophases of Polydisperse Olefin Block Copolymers. *Macromolecules* **2009**, *42* (11), 3788–3794. https://doi.org/10.1021/ma9002819.
(44) Schmitt, A. L.; Mahanthappa, M. K. Polydispersity-Driven Shift in the Lamellar Mesophase Composition Window of PEO-PB-PEO Triblock Copolymers. *Soft Matter* **2012**, *8* (7), 2294–2303. https://doi.org/10.1039/c2sm07041c.
(45) Clair, C.; Lallam, A.; Rosenthal, M.; Sztucki, M.; Vatankhah-Varnosfaderani, M.; Keith, A. N.; Cong, Y.; Liang, H.; Dobrynin, A. V.; Sheiko, S. S.; Ivanov, D. A. Strained Bottlebrushes in Super-Soft Physical Networks. *ACS Macro Lett.* **2019**, *8* (5), 530–534. https://doi.org/10.1021/acsmacrolett.9b00106.
(46) Matsen, M. W.; Bates, F. S. Origins of Complex Self-Assembly in Block Copolymers. *Macromolecules* **1996**, *29* (23), 7641–7644. https://doi.org/10.1021/ma960744q.
(47) Nian, S.; Lian, H.; Gong, Z.; Zhernenkov, M.; Qin, J.; Cai, L. H. Molecular Architecture Directs Linear-Bottlebrush-Linear Triblock Copolymers to Self-Assemble to Soft

(48) Cai, L.-H. Molecular Understanding for Large Deformations of Soft Bottlebrush Polymer Networks. *Soft Matter* **2020**, *16*, 6259–6264. https://doi.org/10.1039/D0SM00759E.

(49) Zhernenkov, M.; Canestrari, N.; Chubar, O.; DiMasi, E. Soft Matter Interfaces Beamline at NSLS-II: Geometrical Ray-Tracing vs. Wavefront Propagation Simulations. In *Advances in Computational Methods for X-Ray Optics III*; Sanchez del Rio, M., Chubar, O., Eds.; International Society for Optics and Photonics, 2014; Vol. 9209, p 92090G. https://doi.org/10.1117/12.2060889.

(50) Pandolfi, R. J.; Allan, D. B.; Arenholz, E.; Barroso-Luque, L.; Campbell, S. I.; Caswell, T. A.; Blair, A.; De Carlo, F.; Fackler, S.; Fournier, A. P.; Freychet, G.; Fukuto, M.; Gürsoy, D.; Jiang, Z.; Krishnan, H.; Kumar, D.; Kline, R. J.; Li, R.; Liman, C.; Marchesini, S.; Mehta, A.; N'Diaye, A. T.; Parkinson, D. Y.; Parks, H.; Pellouchoud, L. A.; Perciano, T.; Ren, F.; Sahoo, S.; Strzalka, J.; Sunday, D.; Tassone, C. J.; Ushizima, D.; Venkatakrishnan, S.; Yager, K. G.; Zwart, P.; Sethian, J. A.; Hexemer, A. Xi-Cam: A Versatile Interface for Data Visualization and Analysis. *J. Synchrotron Radiat.* **2018**, *25* (4), 1261–1270. https://doi.org/10.1107/S1600577518005787.

(51) Kremer, J. R.; Mastronarde, D. N.; McIntosh, J. R. Computer Visualization of Three-Dimensional Image Data Using IMOD. *J. Struct. Biol.* **1996**, *116* (1), 71–76. https://doi.org/10.1006/jsbi.1996.0013.

(52) Gilbert, P. F. The Reconstruction of a Three-Dimensional Structure from Projections and Its Application to Electron Microscopy. II. Direct Methods. *Proc. R. Soc. London. Ser. B. Biol. Sci.* **1972**, *182* (66), 89–102. https://doi.org/10.1098/rspb.1972.0068.

For table of content only

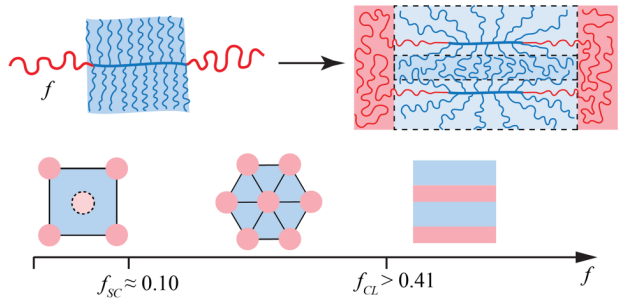

*Supporting Information*

# Anomalous Self-assembly of Architecturally Semiflexible Block Copolymers

Shifeng Nian,[1] Zhouhao Fan[2], Guillaume Freychet,[7] Mikhail Zhernenkov,[7] Stefanie Redemann,[4,5,6], Li-Heng Cai[1,2,3,*]

[1]Soft Biomatter Laboratory, Department of Materials Science and Engineering, University of Virginia, Charlottesville, VA 22904, USA
[2]Department of Chemical Engineering, UVA
[3]Department of Biomedical Engineering, UVA
[4]Department of Molecular Physiology and Biological Physics, UVA
[5]Department of Cell Biology, UVA
[6]Center for Membrane and Cell Physiology, UVA
[7]National Synchrotron Light Source-II, Brookhaven National Laboratory, Upton, NY 11973, USA

*Correspondence to liheng.cai@virginia.edu
ORCID: 0000-0002-6806-0566

**Table of Contents**

# I. Supporting Text

**Characteristic SAXS peaks for different types of microstructures.** We attempt to determine the type of a microstructure by comparing the positions of the characteristic peaks to the allowed reflections for a space group using ratios of $q/q^*$, where $q^*$ is the wavenumber of the primary scattering peak at the lowest wavenumber. This method has been documented in literature for block copolymer self-assembly. Here we briefly summarize the results relevant to this study.

The primary peak is associated with the largest lattice spacing for Bragg's diffraction to occur:
$$d^* = \frac{2\pi}{q^*} \tag{S1}$$

*Body centered cubic (BCC) sphere.* For a cubic system, the lattice spacing is
$$d_{hkl} = \frac{a}{\sqrt{h^2+k^2+l^2}} \tag{S2}$$
in which $a$ is the lattice constant of the cubic crystal, and $h$, $k$, and $l$ are the Miller indices for the Bragg plane. For BCC, the allowable reflections occur when the value of $h + k + l$ is even:
$$q_{hkl} = \frac{2\pi}{a}\sqrt{h^2+k^2+l^2}, \text{ for } h+k+l = \text{even} \tag{S3}$$
Therefore, the primary scattering peak is
$$q^* = q_{110} = \frac{2\pi}{a/\sqrt{2}}, \text{ and } a = \sqrt{2}d^* \tag{S4}$$
and the allowable reflections are
$$q/q^* = \sqrt{1}, \sqrt{2}, \sqrt{3}, \sqrt{4}, \sqrt{5}, \dots \tag{S5}$$

*Hexagonal cylinder (Hex C).* For a two-dimensional hexagonal symmetry, the Bragg reflections correspond to
$$q_{hk} = \frac{2\pi}{a\sqrt{3}/2}\sqrt{h^2+k^2+hk} \tag{S6}$$
in which $a$ is the lattice constant or the nearest neighbor distance. The first seven allowable reflections are:
$$q/q^* = \sqrt{1}, \sqrt{3}, \sqrt{4}, \sqrt{7}, \sqrt{9}, \sqrt{12}, \sqrt{13}, \sqrt{16}, \sqrt{19}, \sqrt{21}, \dots \tag{S7}$$
and
$$q^* = q_{10} = \frac{2\pi}{a\sqrt{3}/2}, \text{ and } a = \left(\frac{2}{\sqrt{3}}\right)d^* \tag{S8}$$

*Lamellar structure.* For a one-dimensional lamellar structure, the Bragg reflections correspond to
$$q_h = \frac{2\pi}{a}h \tag{S9}$$
in which $a$ is the lattice constant, or the spacing of the lamellar stack. The allow reflections are
$$q/q^* = 1, 2, 3, 4, 5, \dots \tag{S10}$$
and
$$q^* = q_1 = \frac{2\pi}{a}, \text{ and } a = d^* \tag{S11}$$

In our study, it is difficult to resolve the type of microstructure based on the scattering pattern because of relatively broad scattering peaks. Moreover, the peak relation from different microstructures overlaps with each other, as shown by eqs. S5, S7, and S10 respectively for BCC sphere, hexagonal cylinder, and lamellae. To this end, we use TEM characterization to complement SAXS measurements; this allows for determining the type of microstructure unambiguously.

## II. Supporting Figures and Tables

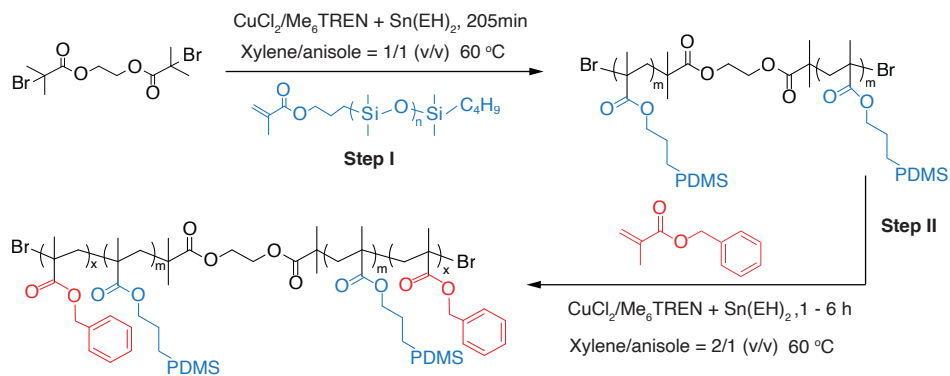

**Figure S1. Scheme of polymer synthesis.**
*Step I:* ARGET ATRP of methacrylate terminated PDMS to synthesize a semiflexible bottlebrush. *Step II:* Synthesis of the linear PBnMA blocks. We control the reaction from 1 to 6 hours to increase the molecular weight of the end blocks.

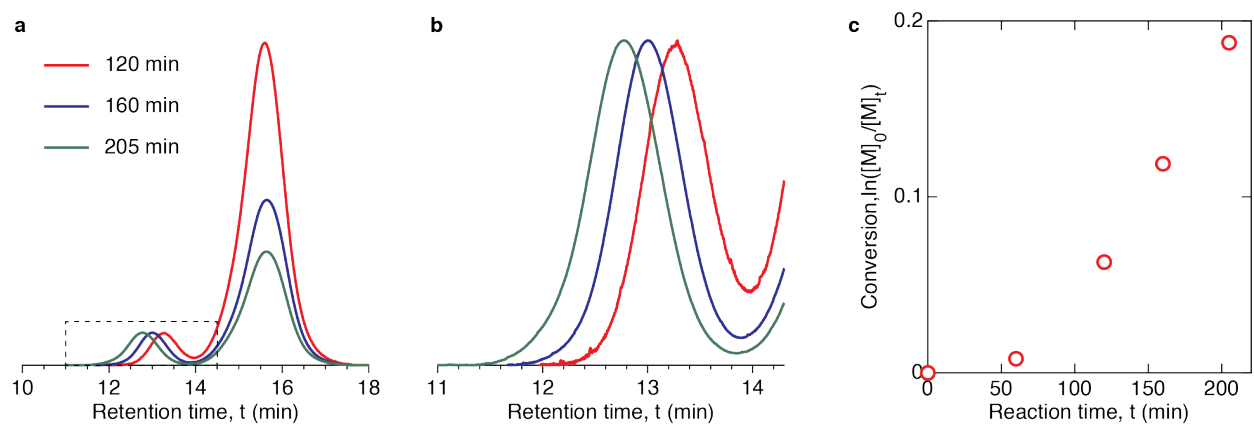

**Figure S2. GPC traces and kinetics of ARGET ATRP of PDMS macromonomers.**
**(a)** GPC traces of raw reaction mixture. **(b)** Zoom-in view of the GPC traces for bbPDMS in **(a)**. **(c)** Kinetics of the polymerization.

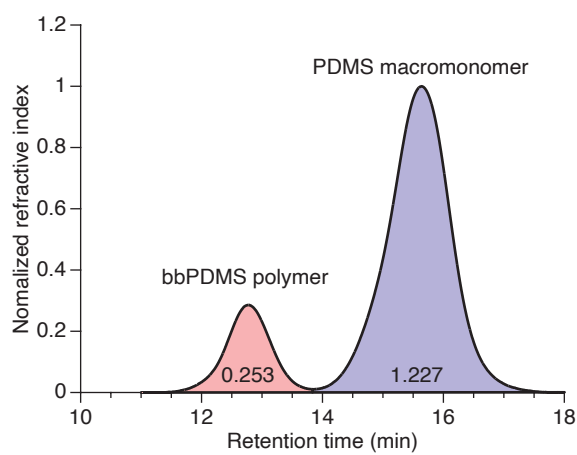

**Figure S3. GPC trace of product mixture from ARGET ATRP of PDMS macromonomer.**
The peak with a shorter retention time corresponds to the bbPDMS polymer, whereas the peak with a longer retention time corresponds to the PDMS macromonomer. The area under each peak represents the mass of the molecule. Based on the area associated with the bbPDMS polymer, 0.2533, and that associated with the PDMS macromonomer, 1.2271, it can be determined the conversion of macromonomer: 0.2533×100%/(0.2533+1.2271)=17.1%. Thus, the DP of macromonomer is 300 × 17.1% = 51, consistent with that obtained from NMR (see **SI NMR spectra**)

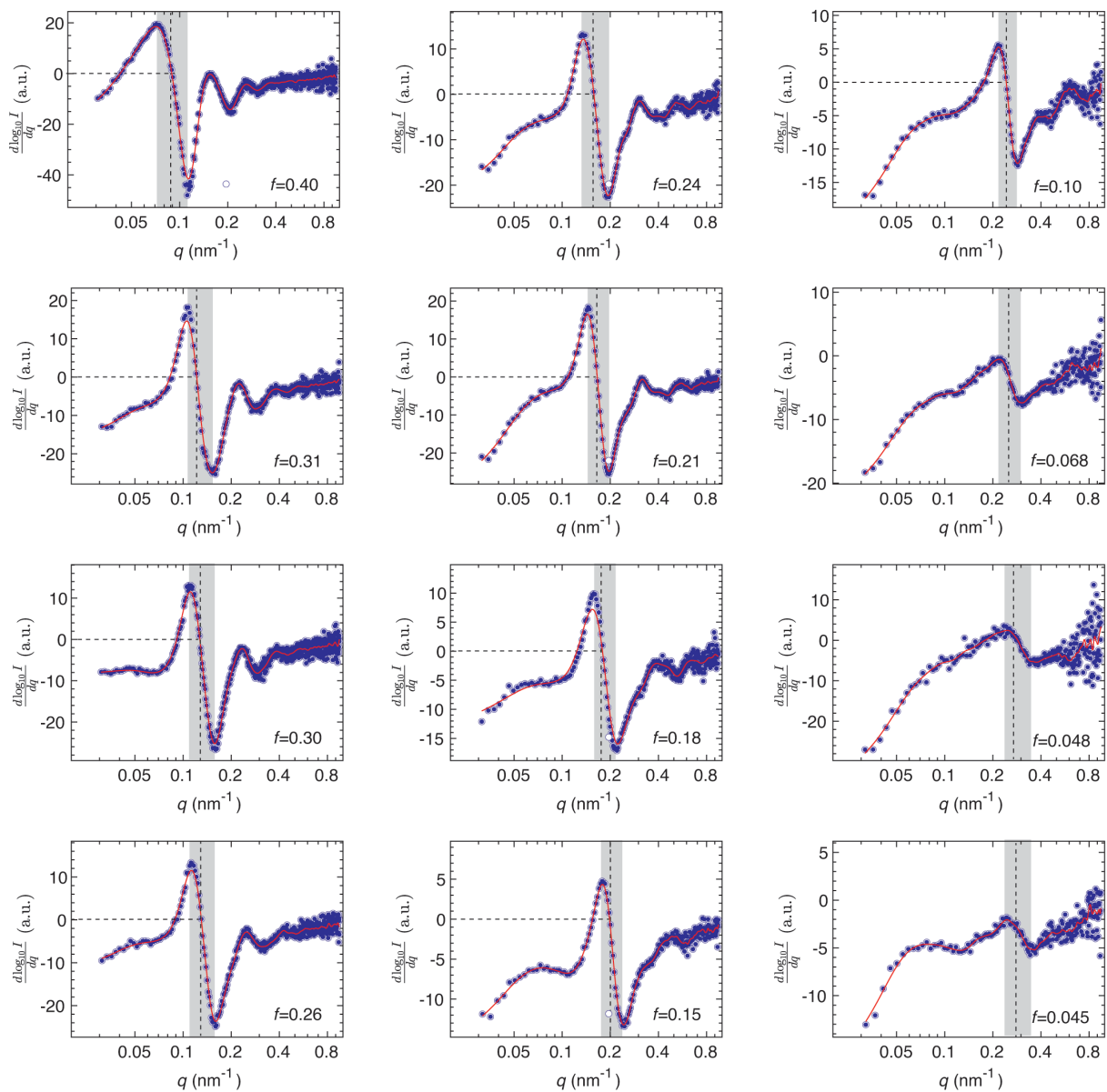

**Figure S4. Determination of characteristic peaks of SAXS measurements.**
The dependence of derivative $d\,(\log_{10} I)/dq$ on $q$ for triblock copolymers with the same bottlebrush (batch 1 in Table 1). Solid red lines are the spline fit to the data as the guidance for the eye.

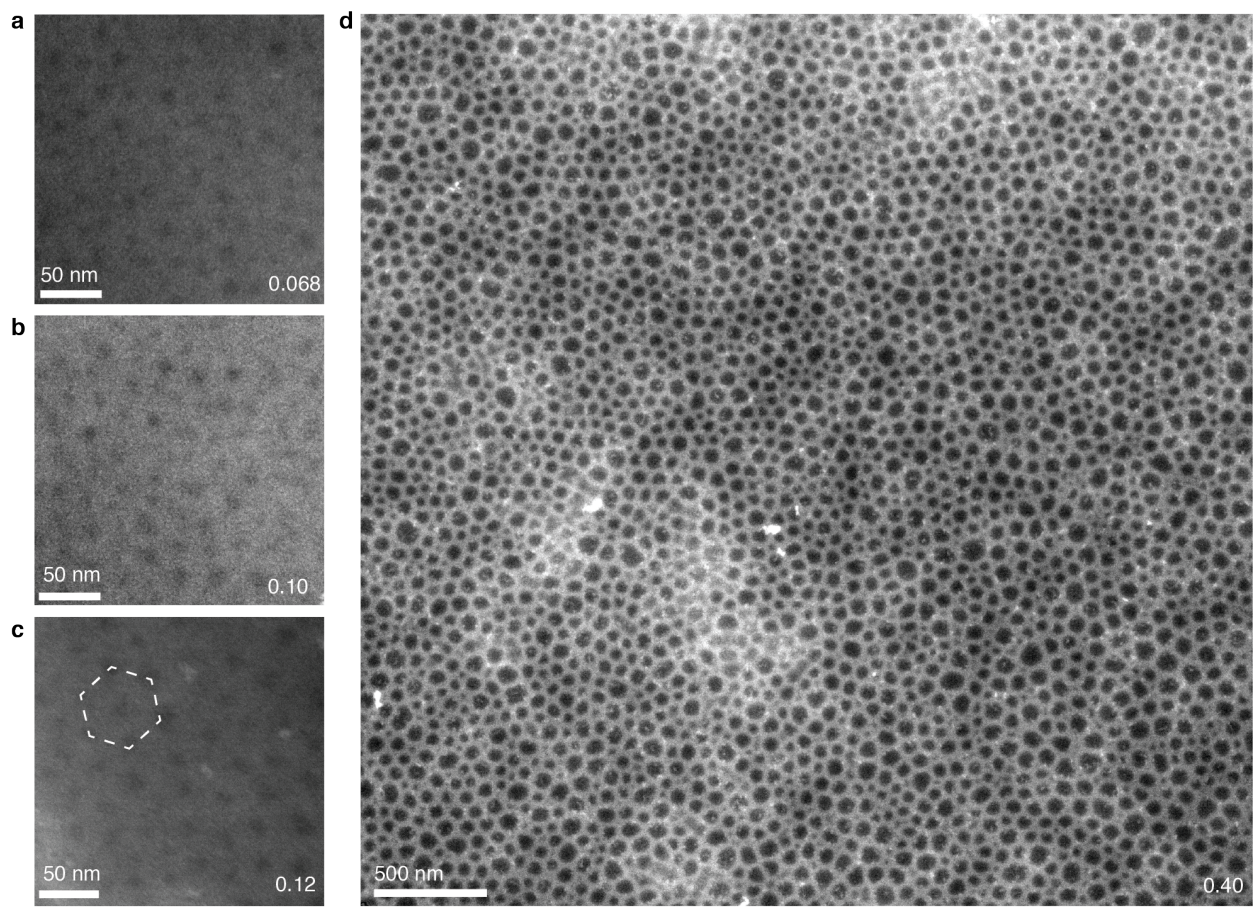

**Figure S5. TEM images of the self-assembled microstructures.**
(**a**) sample *f*=0.067, batch 1; (**b**) sample *f*=0.10, batch 1; (**c**) sample *f*=0.12, batch 4; (**d**) sample *f*=0.40, batch 1. Dark regions are domains formed by end linear PBnMA blocks, and white regions are bottlebrush PDMS domains.

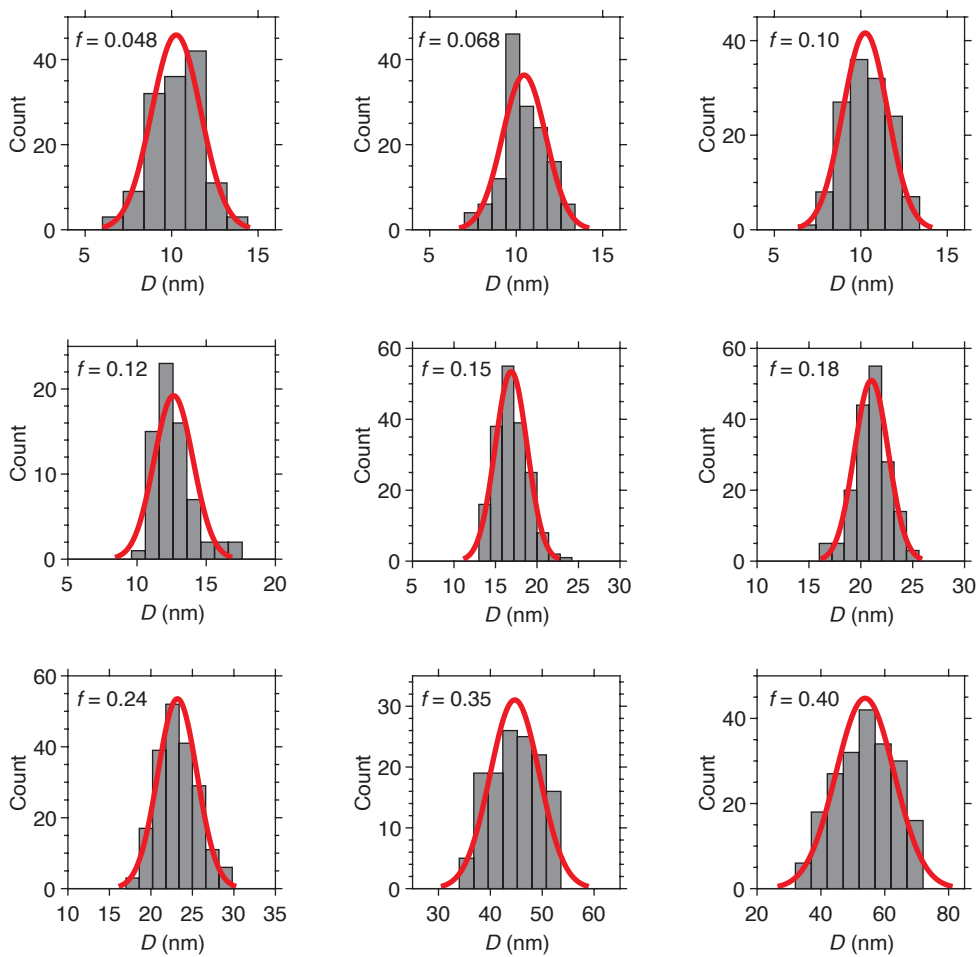

**Figure S6. Domain diameter directly measured from TEM images.**
The values for the diameter, $D$, of each polymer are measured using ImageJ and are listed in **Table 1**.

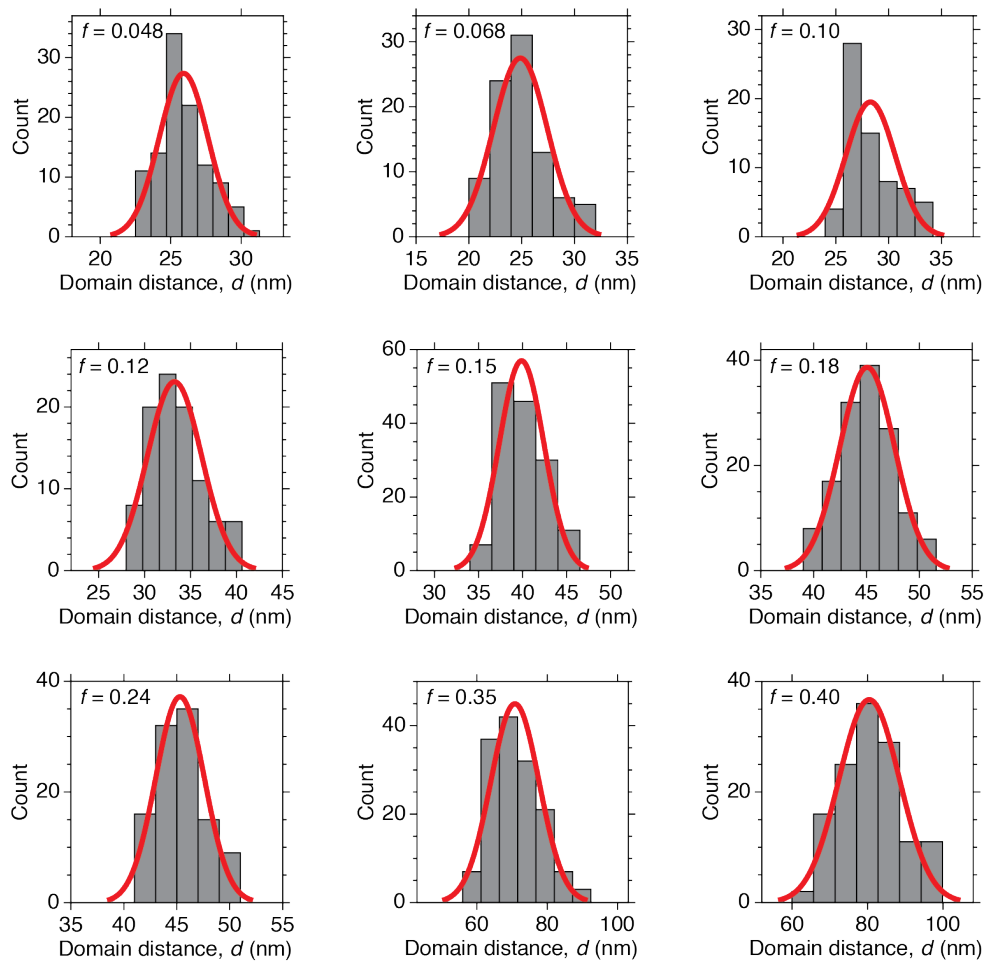

**Figure S7. Domain distance directly measured from TEM images.**
The values for the distance, *d*, of each polymer are measured using ImageJ and are listed in **Table S1**.

**Table S1. Volume fraction of selected samples calculated from TEM and NMR.**

| Sample | Morphology | PBnMA domain distance from TEM (nm) | PBnMA domain diameter from TEM (nm) | $f$ calculated from TEM | $f$ calculated from NMR |
|---|---|---|---|---|---|
| $S_3$ | BCC S | 25.9±1.7 | 10.3±1.4 | 0.066±0.013 | 0.048 |
| $S_4$ | BCC S | 24.8±2.6 | 10.4±1.3 | 0.077±0.024 | 0.068 |
| $S_5$ | Hex C | 28.3±2.3 | 10.2±1.3 | 0.12±0.020 | 0.10 |
| $S_6$ | Hex C | 39.9±2.5 | 16.9±1.9 | 0.16±0.020 | 0.15 |
| $S_7$ | Hex C | 45.1±2.6 | 21.0±1.6 | 0.20±0.023 | 0.18 |
| $S_9$ | Hex C | 45.3±2.3 | 22.6±2.4 | 0.23±0.023 | 0.24 |
| $S_{13}$ | Hex C | 70.8±6.9 | 45.4±5.4 | 0.37±0.072 | 0.35 |
| $S_{14}$ | Hex C | 80.4±8.0 | 53.8±9.2 | 0.41±0.082 | 0.40 |
| $S_{20}$ | Hex C | 33.2±3.0 | 12.6±1.1 | 0.13±0.023 | 0.12 |

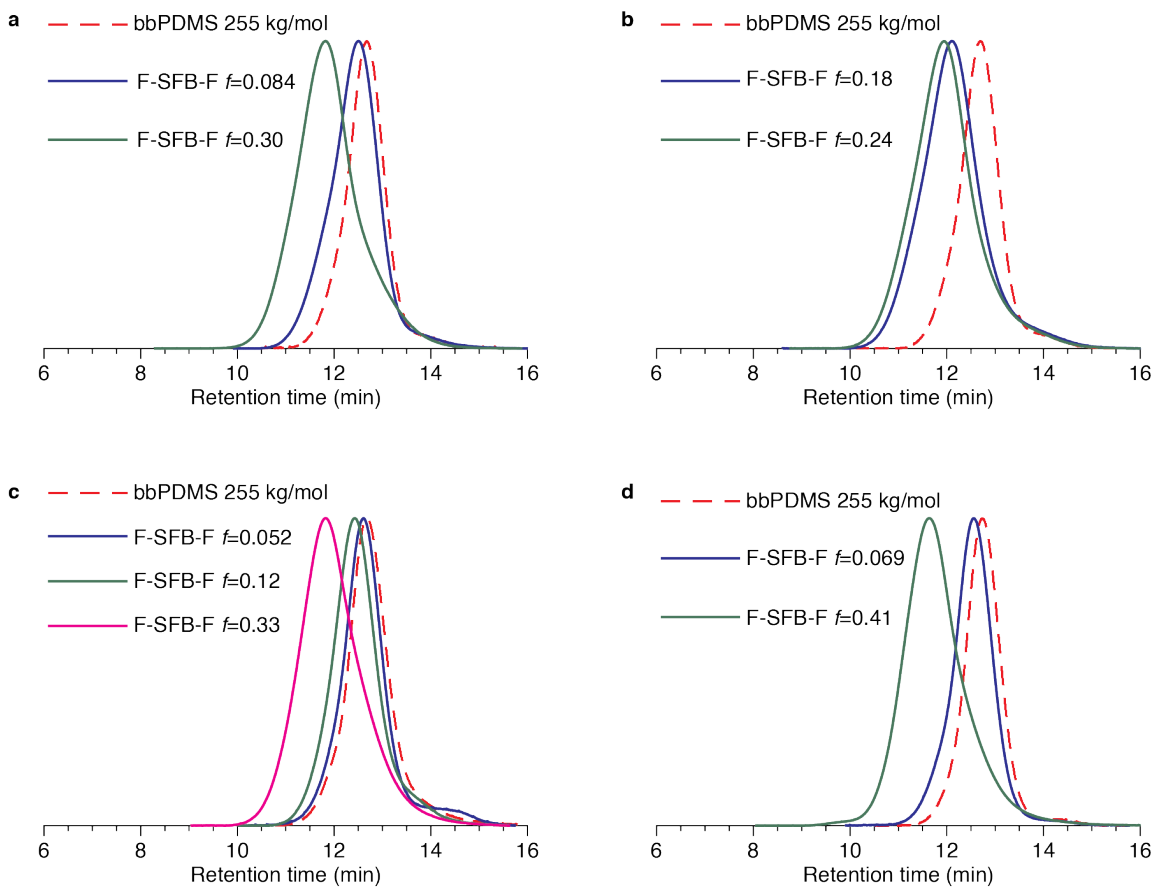

**Figure S8. GPC traces of different batches of F-SFB-F triblock polymers.**
(**a**) Batch 2 with bbPDMS PDI of 1.18. (**b**) Batch 3 with bbPDMS PDI of 1.20. (**c**) Batch 4 with bbPDMS PDI of 1.17. (**d**) Batch 5 with bbPDMS PDI of 1.16.

## III. $^1$H NMR Spectra

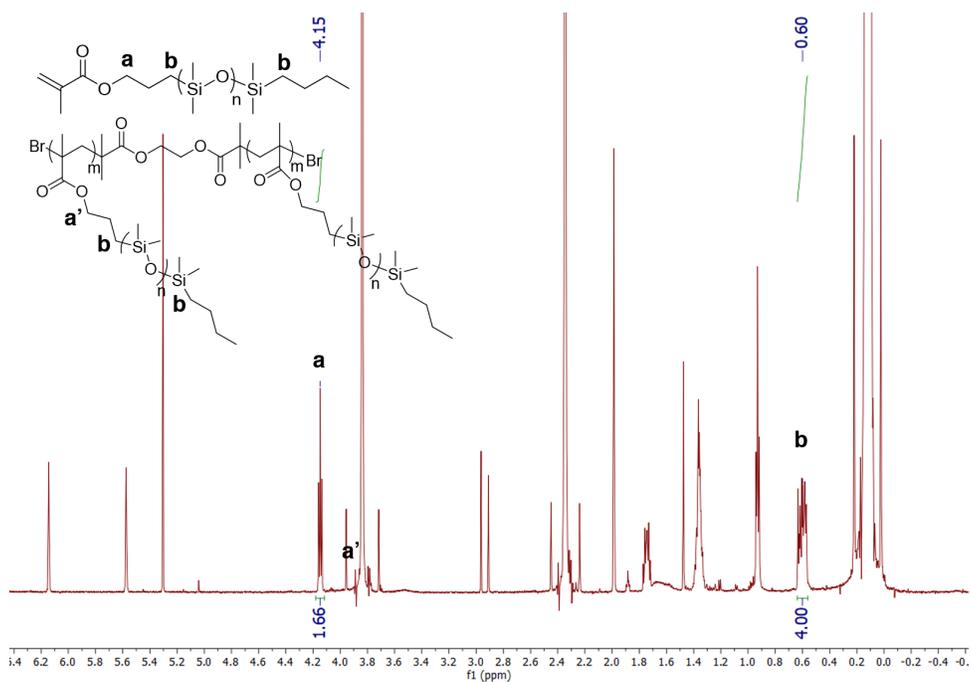

**Batch 1** PDMS macromonomer.
Conversion = [1 - Area(**a**) × 2/Area(**b**)]×100%=17%. DP of macromonomer is 300×17%=51.

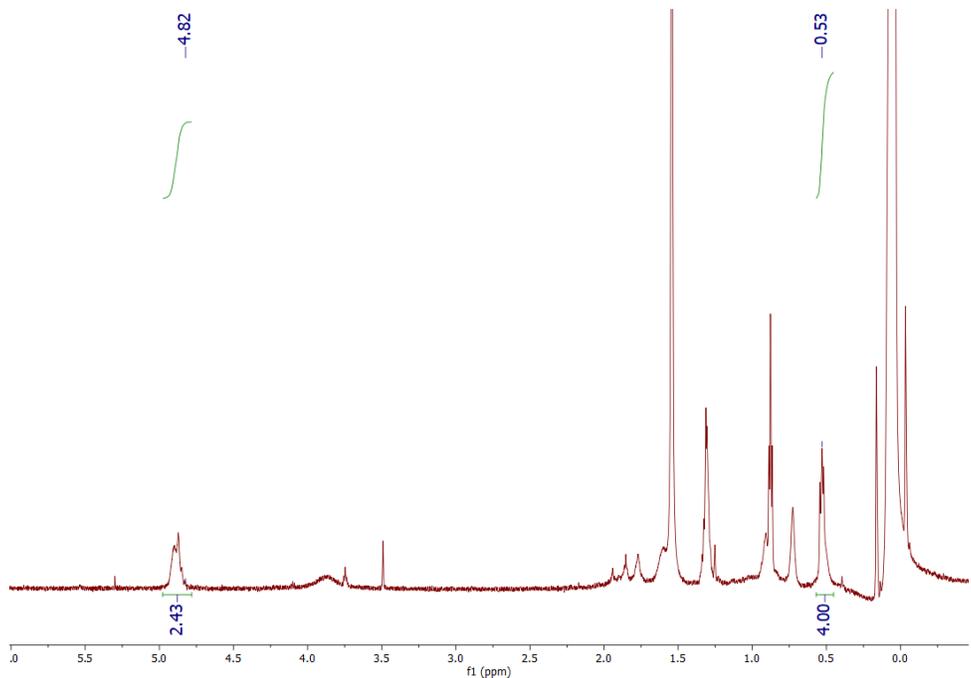

**Batch 1** LBBL polymer BnMA$_{31}$-*b*-PDMS$^5_{51}$-*b*-BnMA$_{31}$.
DP of BnMA = 51×2.43/2 = 62.

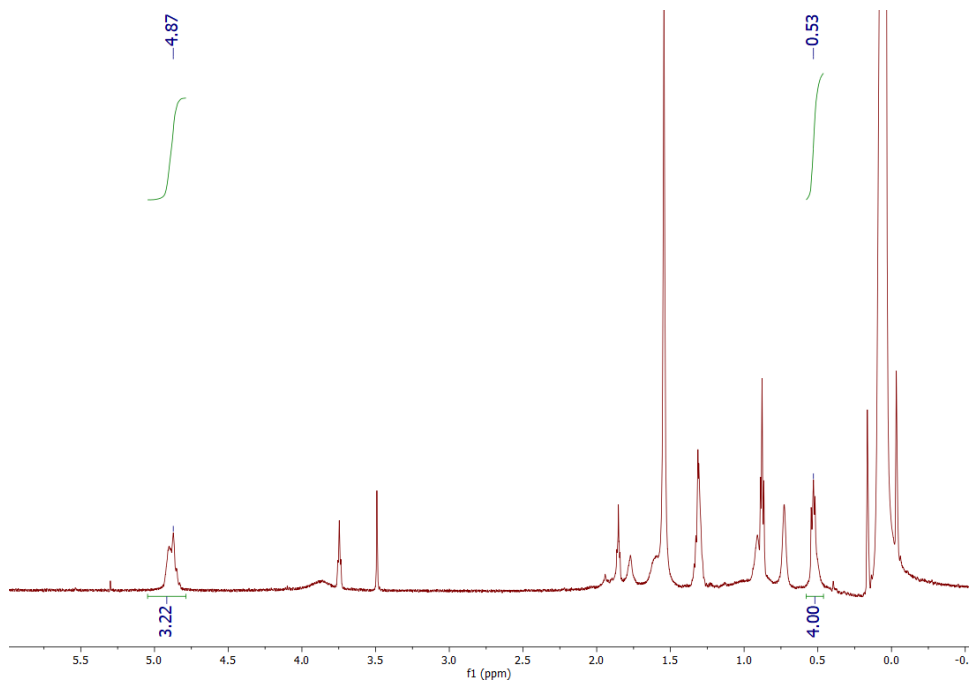

**Batch 1** LBBL polymer BnMA$_{41}$-*b*-PDMS$^5_{51}$-*b*-BnMA$_{41}$.
DP of BnMA = 51×3.22/2 = 82.

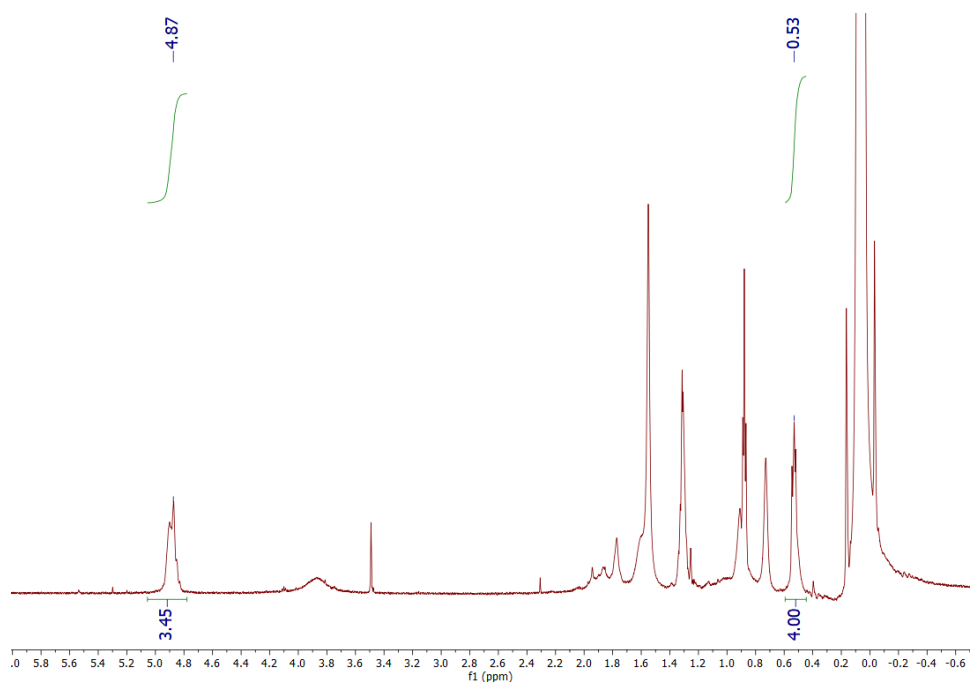

**Batch 1** LBBL polymer BnMA$_{44}$-*b*-PDMS$^5_{51}$-*b*-BnMA$_{44}$.
DP of BnMA = 51×3.45/2 = 88.

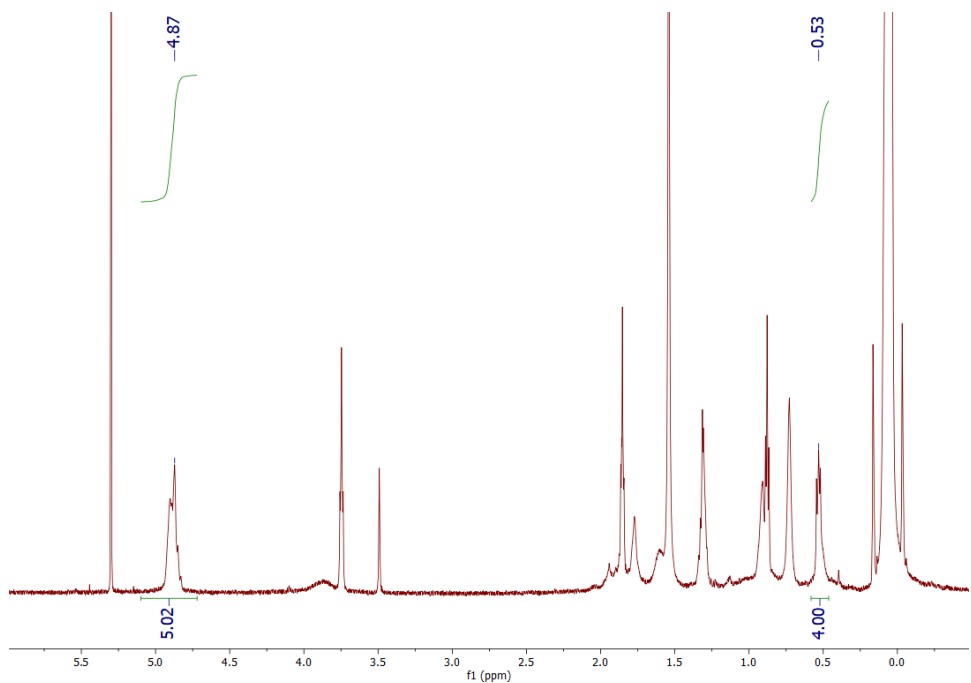

**Batch 1** LBBL polymer BnMA$_{64}$-*b*-PDMS$^{5}_{51}$-*b*-BnMA$_{64}$.
DP of BnMA = 51×5.02/2 = 128.

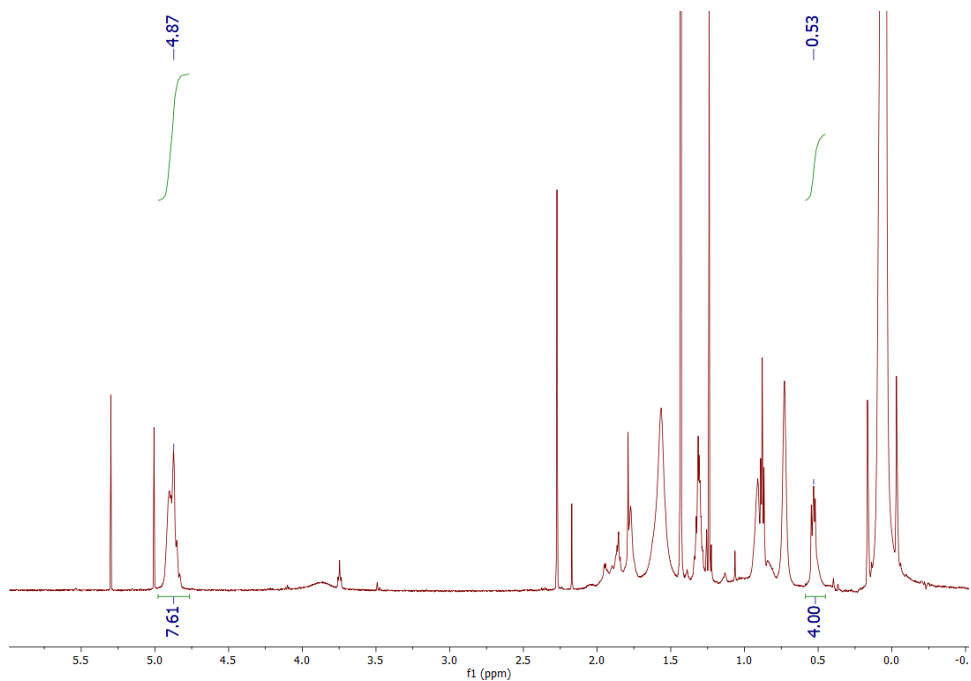

**Batch 1** LBBL polymer BnMA$_{97}$-*b*-PDMS$^{5}_{51}$-*b*-BnMA$_{97}$.
DP of BnMA = 51×7.61/2 = 194.

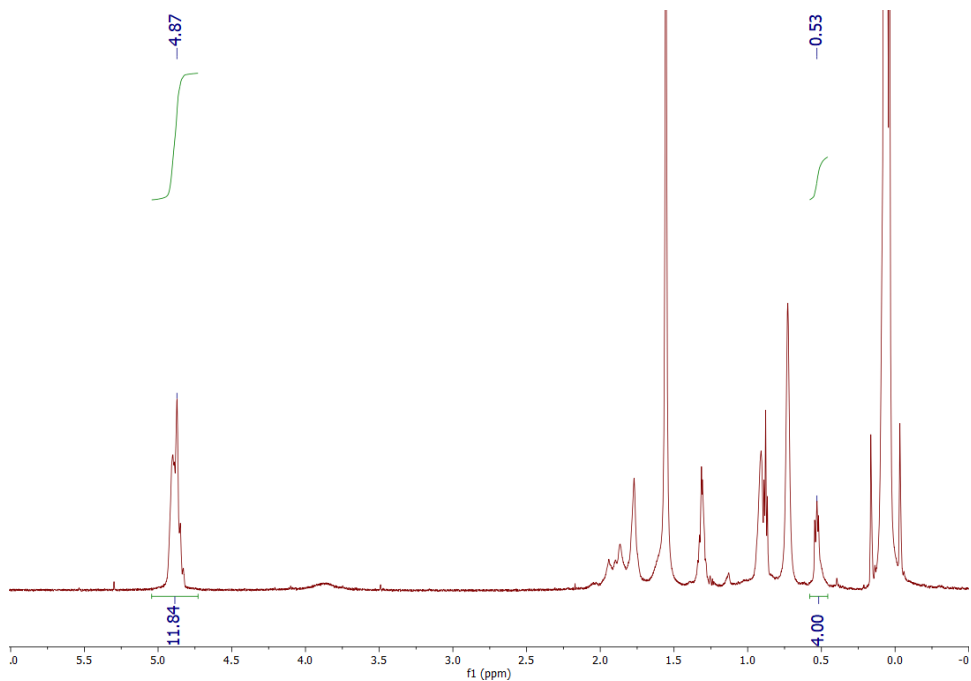

**Batch 1** LBBL polymer BnMA$_{151}$-*b*-PDMS$^5_{51}$-*b*-BnMA$_{151}$.
DP of BnMA = 51×11.84/2 = 302.

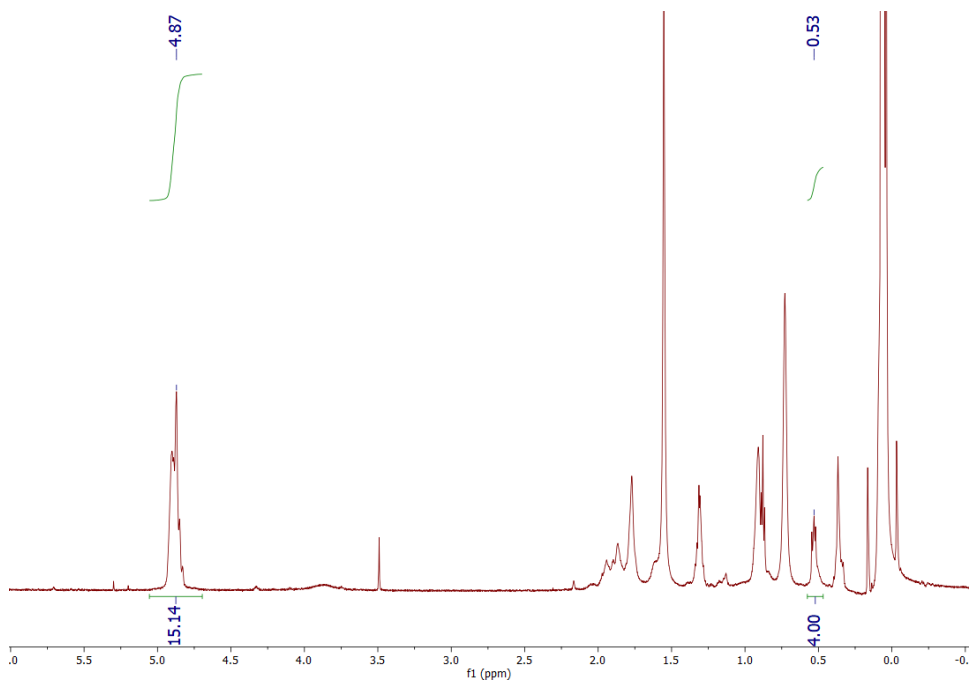

**Batch 1** LBBL polymer BnMA$_{193}$-*b*-PDMS$^5_{51}$-*b*-BnMA$_{193}$.
DP of BnMA = 51×15.14/2 = 386.

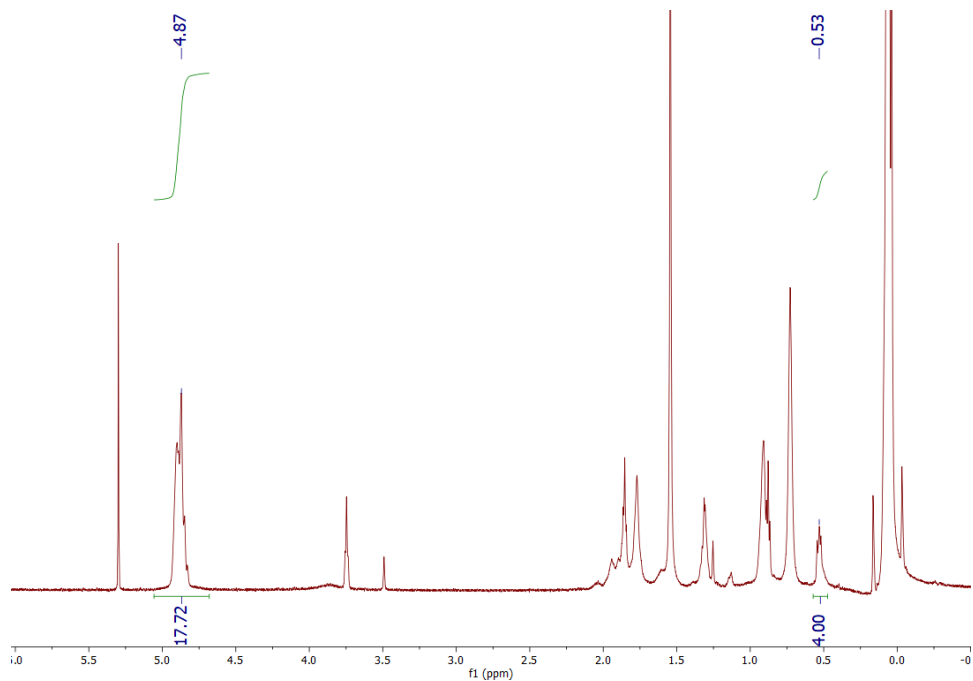

**Batch 1** LBBL polymer BnMA$_{226}$-*b*-PDMS$^5_{51}$-*b*-BnMA$_{226}$.
DP of BnMA = 51×17.72/2 = 452.

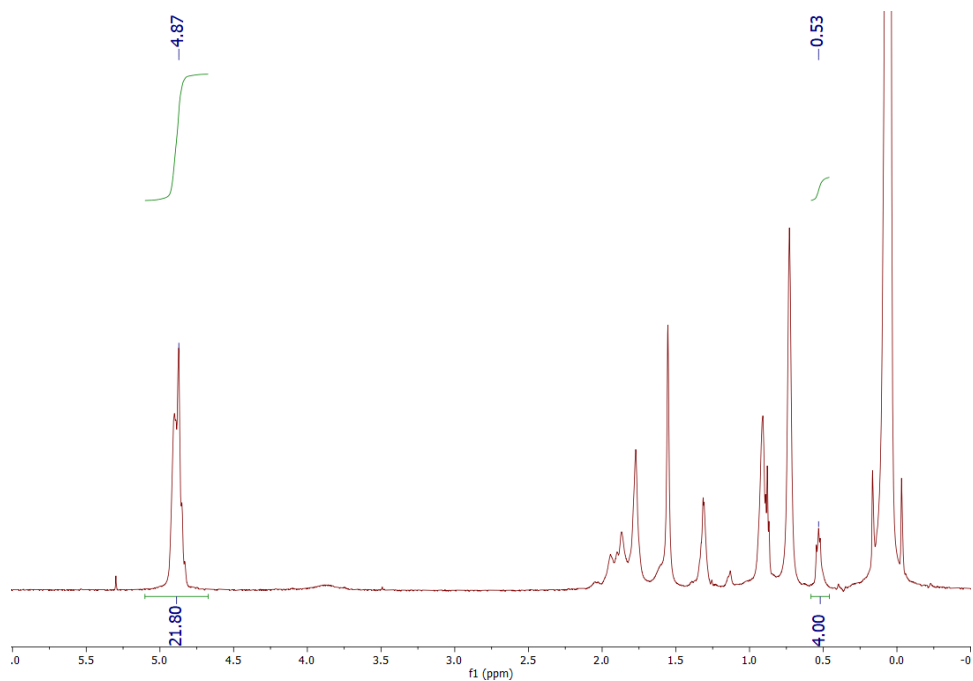

**Batch 1** LBBL polymer BnMA$_{278}$-*b*-PDMS$^5_{51}$-*b*-BnMA$_{278}$.
DP of BnMA = 51×21.80/2 = 556.

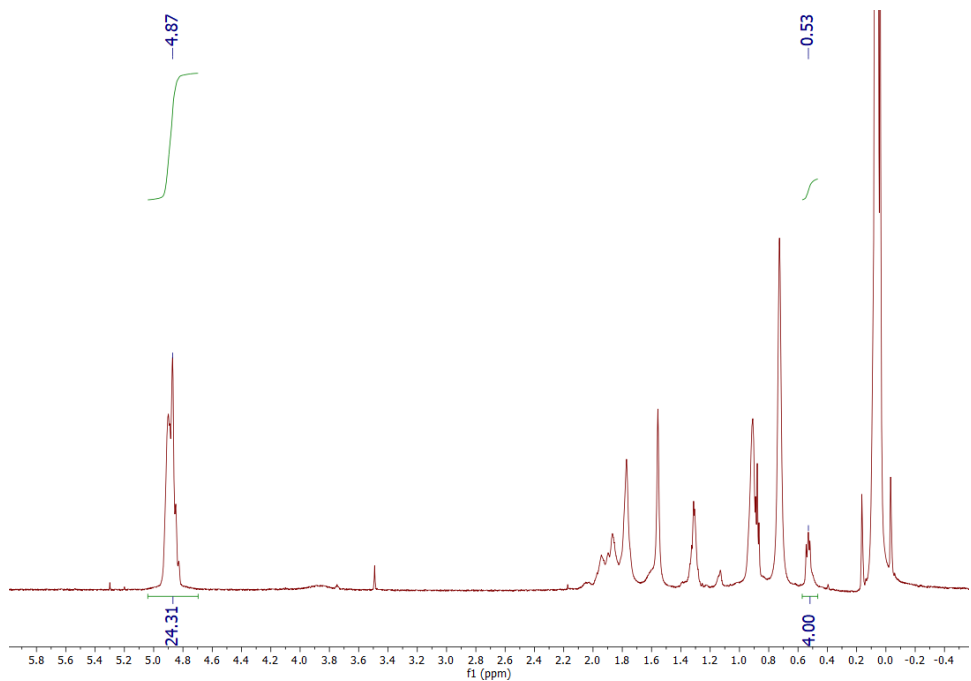

**Batch 1** LBBL polymer BnMA$_{310}$-*b*-PDMS$^5_{51}$-*b*-BnMA$_{310}$.
DP of BnMA = 51×24.31/2 = 620.

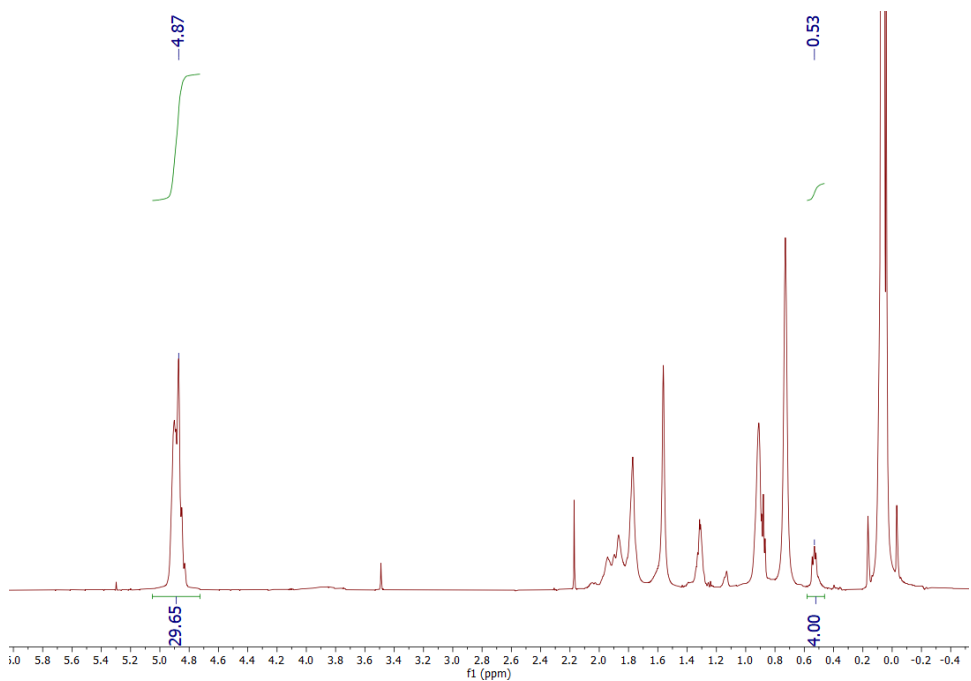

**Batch 1** LBBL polymer BnMA$_{378}$-*b*-PDMS$^5_{51}$-*b*-BnMA$_{378}$.
DP of BnMA = 51×29.65/2 = 756.

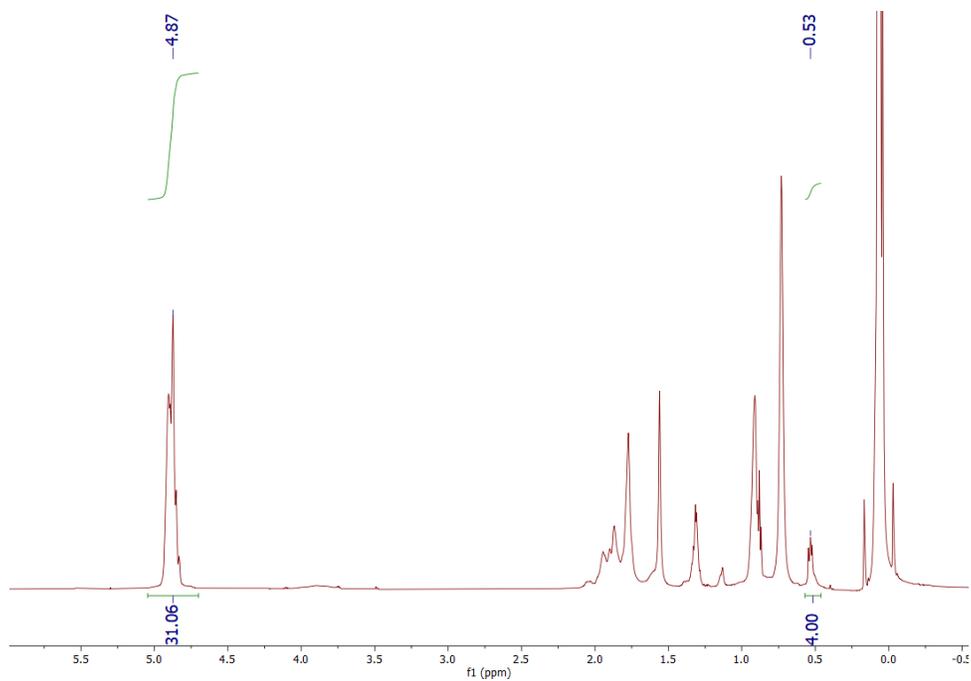

**Batch 1** LBBL polymer BnMA$_{396}$-*b*-PDMS$^5_{51}$-*b*-BnMA$_{396}$.
DP of BnMA = 51×31.06/2 = 792.

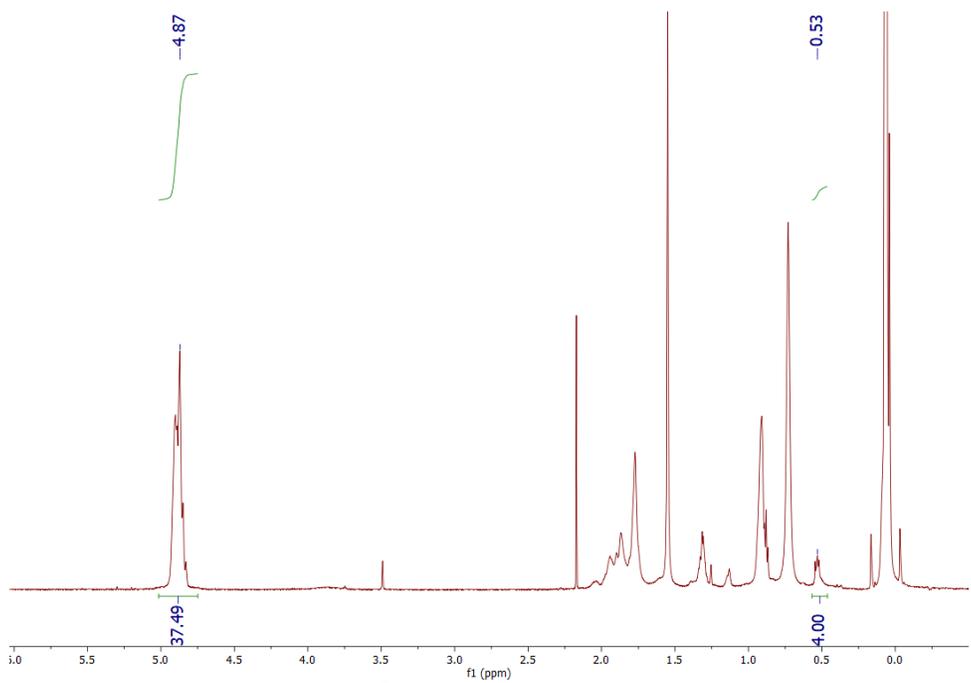

**Batch 1** LBBL polymer BnMA$_{478}$-*b*-PDMS$^5_{51}$-*b*-BnMA$_{478}$.
DP of BnMA = 51×37.49/2 = 956.

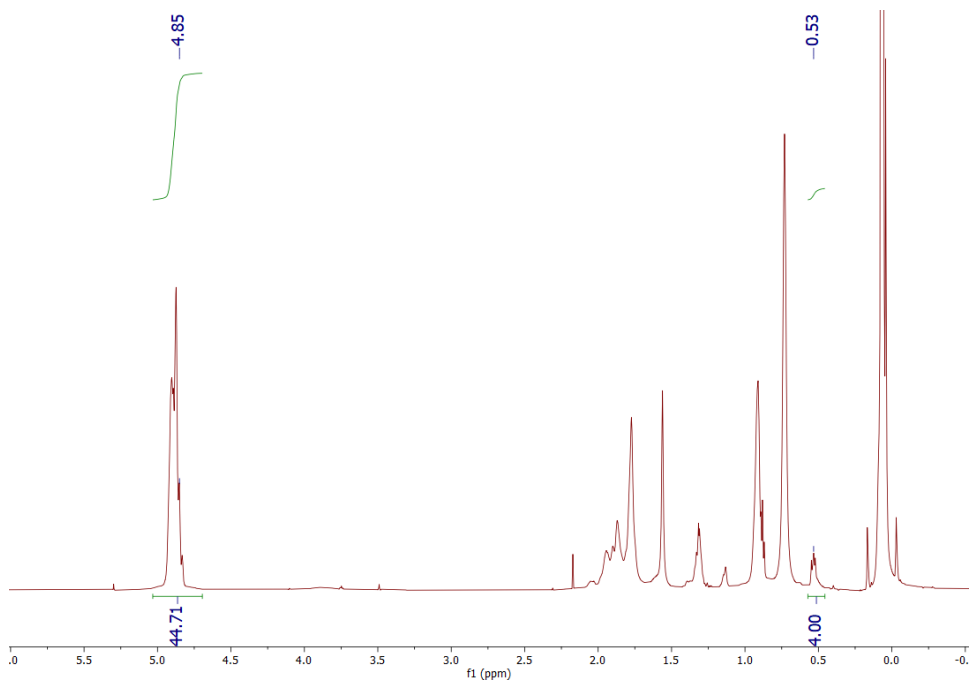

**Batch 1** LBBL polymer $BnMA_{570}$-$b$-$PDMS^5_{51}$-$b$-$BnMA_{570}$.
DP of BnMA = 51×44.71/2 = 1140.

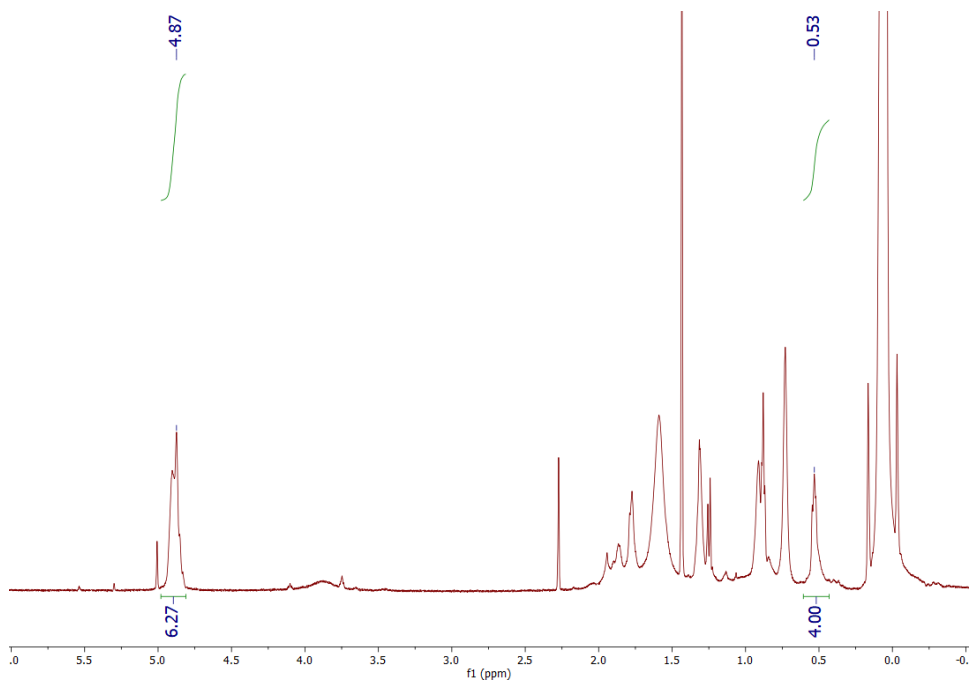

**Batch 2** LBBL polymer $BnMA_{80}$-$b$-$PDMS^5_{51}$-$b$-$BnMA_{80}$.
DP of BnMA = 51×6.27/2 = 160.

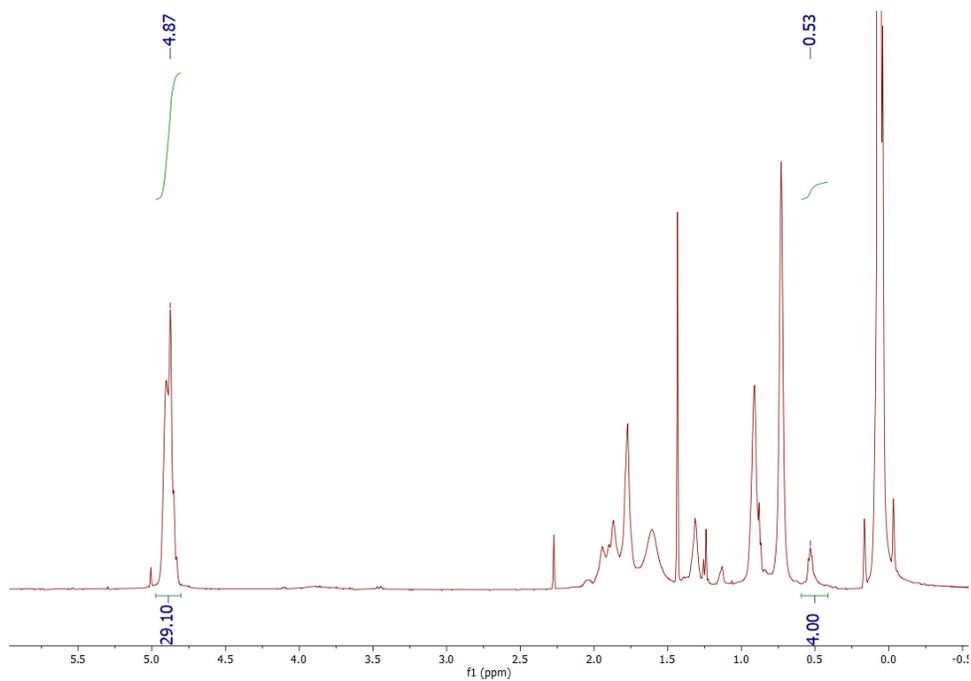

**Batch 2** LBBL polymer BnMA$_{371}$-b-PDMS$^5_{51}$-b-BnMA$_{371}$.
DP of BnMA = 51×29.1/2 = 742.

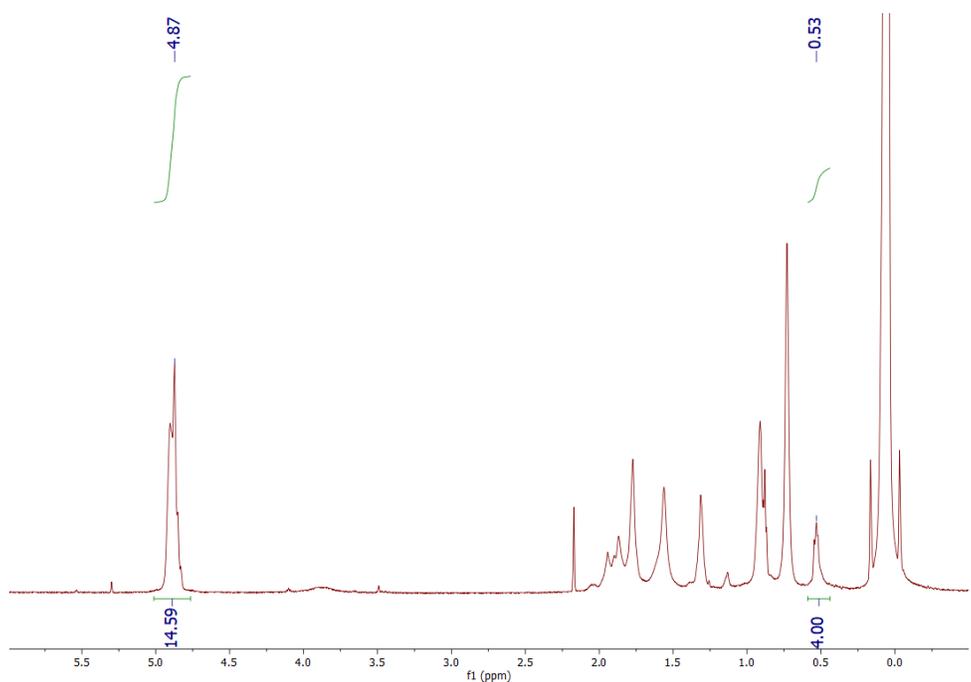

**Batch 3** LBBL polymer BnMA$_{186}$-b-PDMS$^5_{51}$-b-BnMA$_{186}$.
DP of BnMA = 51×14.59/2 = 372.

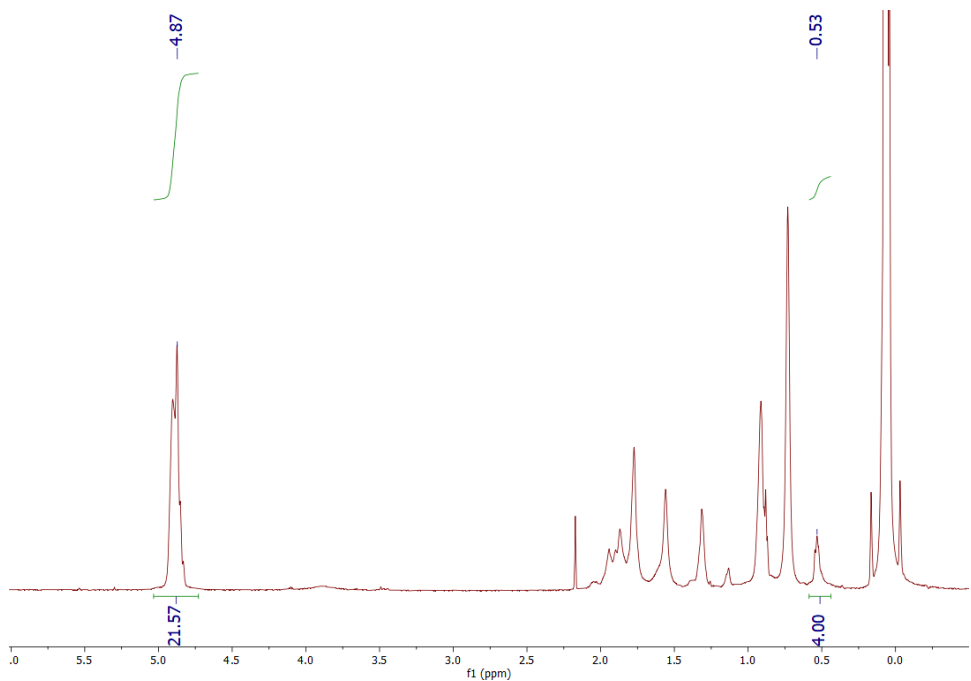

**Batch 3** LBBL polymer BnMA$_{275}$-*b*-PDMS$^5_{51}$-*b*-BnMA$_{275}$.
DP of BnMA = 51×21.57/2 = 550.

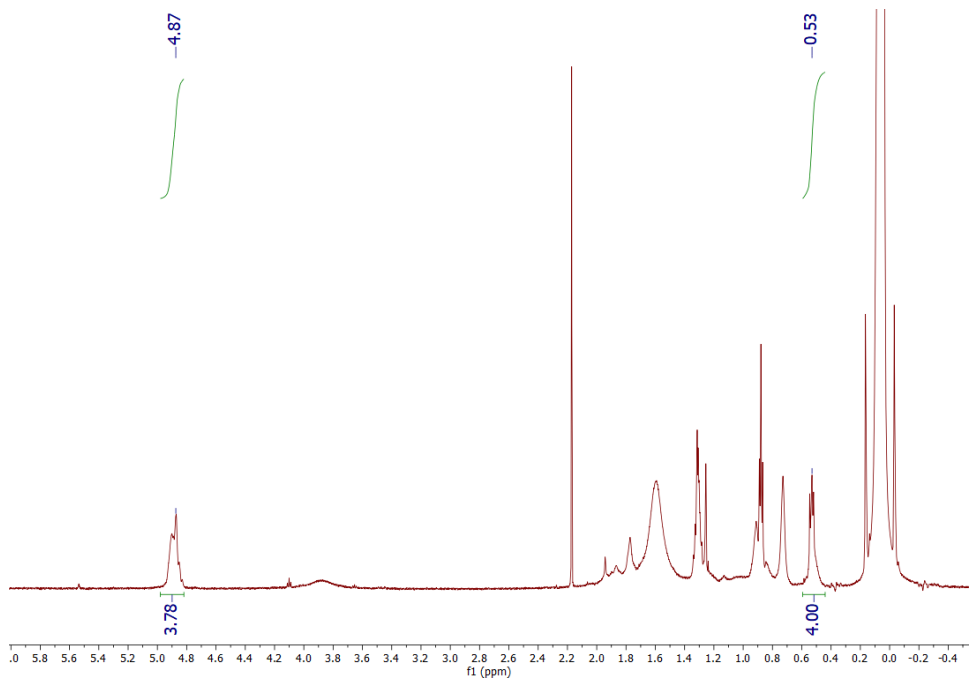

**Batch 4** LBBL polymer BnMA$_{48}$-*b*-PDMS$^5_{51}$-*b*-BnMA$_{48}$.
DP of BnMA = 51×3.78/2 = 96.

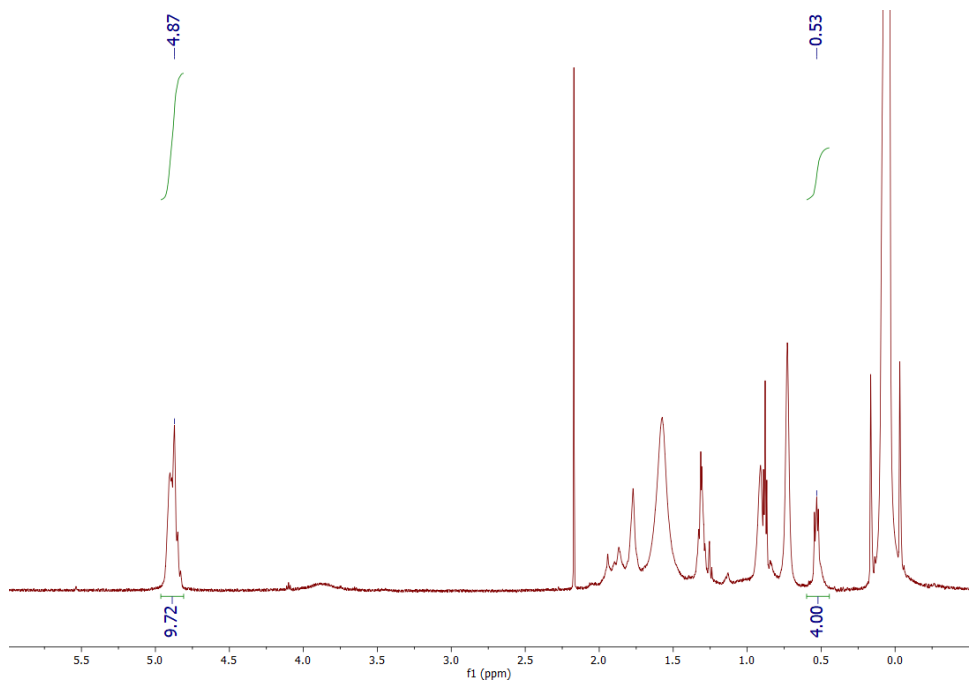

**Batch 4** LBBL polymer BnMA$_{124}$-b-PDMS$^5_{51}$-b-BnMA$_{124}$.
DP of BnMA = 51×9.72/2 = 248.

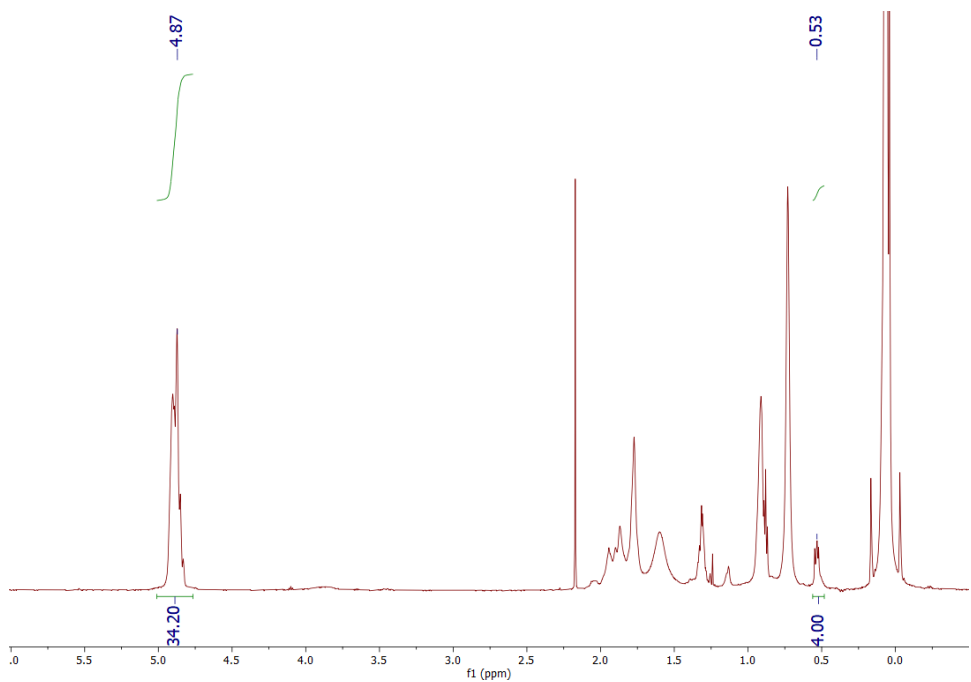

**Batch 4** LBBL polymer BnMA$_{436}$-b-PDMS$^5_{51}$-b-BnMA$_{436}$.
DP of BnMA = 51×34.20/2 = 872.

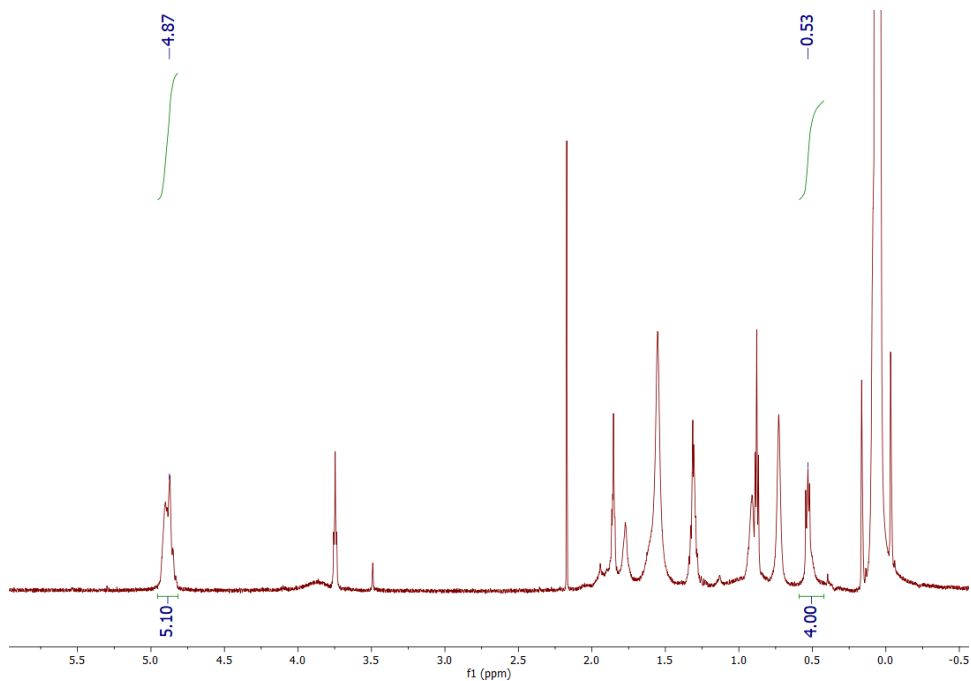

**Batch 5** LBBL polymer BnMA$_{65}$-$b$-PDMS$^5_{51}$-$b$-BnMA$_{65}$.
DP of BnMA = 51×5.10/2 = 130.

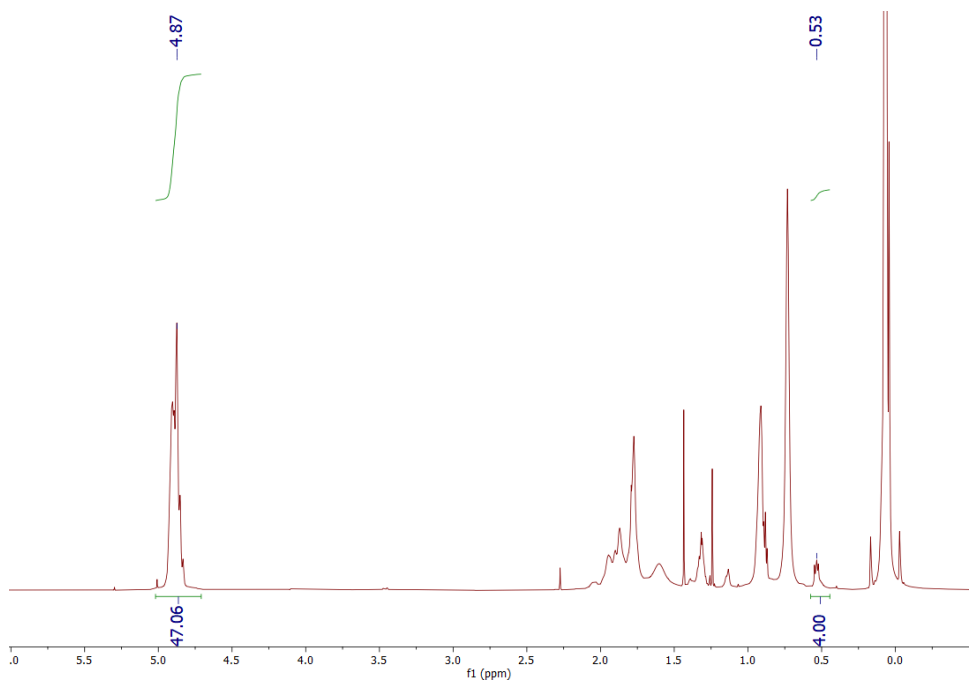

**Batch 5** LBBL polymer BnMA$_{600}$-$b$-PDMS$^5_{51}$-$b$-BnMA$_{600}$.
DP of BnMA = 51×47.06/2 = 1200.

22